\documentclass[final,3p,times]{elsarticle}

\usepackage{amssymb}
\usepackage{amsmath}

\usepackage{lineno}

\newcommand*\patchAmsMathEnvironmentForLineno[1]{
  \expandafter\let\csname old#1\expandafter\endcsname\csname #1\endcsname
  \expandafter\let\csname oldend#1\expandafter\endcsname\csname end#1\endcsname
  \renewenvironment{#1}
     {\linenomath\csname old#1\endcsname}
     {\csname oldend#1\endcsname\endlinenomath}}
\newcommand*\patchBothAmsMathEnvironmentsForLineno[1]{
  \patchAmsMathEnvironmentForLineno{#1}
  \patchAmsMathEnvironmentForLineno{#1*}}
\AtBeginDocument{
\patchBothAmsMathEnvironmentsForLineno{equation}
\patchBothAmsMathEnvironmentsForLineno{align}
\patchBothAmsMathEnvironmentsForLineno{flalign}
\patchBothAmsMathEnvironmentsForLineno{alignat}
\patchBothAmsMathEnvironmentsForLineno{gather}
\patchBothAmsMathEnvironmentsForLineno{multline}
}

\usepackage{xcolor} 
\usepackage{ulem} 
\usepackage{cancel} 
\usepackage{bm}
\usepackage{hyperref}
\biboptions{numbers,sort&compress} 

\begin{document}

\begin{frontmatter}



\title{Enhanced numerical approaches for modeling insoluble surfactants in two-phase flows with the diffuse-interface method}


\author[isct]{Shu Yamashita} 
\author[isct]{Shintaro Matsushita} 
\author[isct]{Tetsuya Suekane} 

\affiliation[isct]{organization={Institute of Science Tokyo, School of Engineering},
            addressline={2-12-1, Ookayama, Meguro-ku}, 
            city={Tokyo},
            postcode={152-8550}, 
            country={Japan}}

\begin{abstract}
Surfactants reside at the interface of two-phase flows and significantly influence the flow dynamics. 
Numerical simulations are essential for a comprehensive understanding of such surfactant-laden flows and require a method that can accurately simulate surfactant transport along the interface.
In this study, we focus on interfacial transport models for insoluble surfactants based on the diffuse-interface method and propose two approaches to improve their accuracy: (a) adopting a formulation that avoids the spatial derivatives of variables with sharp gradients and (b) allowing the width of the delta function to be specified independently of the interface width. 
These approaches are simple and practical in that they do not lead to significant increases in computational cost, implementation complexity, or degradation of interface-capturing accuracy. 
Moreover, they preserve the discrete conservation of both fluid and surfactant mass.
We conduct a series of numerical tests to demonstrate the effectiveness of the proposed approaches. 
Finally, we present a challenging test case that is difficult to solve accurately and has not been previously discussed.
We expect this case to serve as a valuable benchmark for evaluating and comparing the performances of various methods proposed in the literature.
\end{abstract}




\end{frontmatter}


\section{Introduction}

Understanding the characteristics of two-phase flows containing surfactants is crucial for a wide range of practical applications. 
Surfactants are adsorbed at the interface and significantly influence two-phase flows by locally reducing the surface tension and inducing Marangoni effects. 
These phenomena have been utilized in various fields, including detergents~\cite{Bajpai_2007}, the food industry~\cite{NITSCHKE2007252}, enhanced oil recovery~\cite{Belhaj2020,MASSARWEH20203150}, soil remediation~\cite{TRELLU2016149}, and microfluidics~\cite{Baret2012review,Anna2016review,KOVALCHUK2023102844}. 
To further advance these applications, a deeper understanding of two-phase flows in the presence of surfactants is necessary.

Numerical simulations have been widely used to investigate two-phase flows involving surfactants~\cite{Batchvarov_2021,Constante-Amores_2021,Kalli_2023,Constante-Amores_2023,Pico_2024,Eshima2025}. 
They enable the evaluation of quantities that are difficult to measure experimentally, such as interfacial concentration fields~\cite{Dong_2019} and local force distribution. 
Moreover, the ability to freely adjust the parameters makes it straightforward to examine the effects of specific variables on the flow. 
Consequently, numerical simulations have become invaluable tools for analyzing surfactant-laden two-phase flows and have attracted significant research attention in recent years.

The accurate modeling of surfactant transport is critical for reliably simulating two-phase flows containing surfactants.
For the insoluble surfactants considered in this study, the transport of the surfactant is governed by an advection–diffusion equation on a moving interface. 
To achieve high accuracy in such simulations, various models have been developed based on different interface-capturing and interface-tracking methods.
For example, models have been proposed based on the volume-of-fluid method~\cite{RENARDY200249,JAMES2004685}, the level-set method~\cite{Xu2003,XU2006590,XU20125897,XU201471,XU2018336}, and the front-tracking method~\cite{MURADOGLU20082238,MURADOGLU2014737,SHIN2018}.
However, these methods face challenges in simultaneously ensuring mass conservation of both the fluid and surfactants while maintaining algorithmic simplicity.

This study focuses on modeling surfactant transport using the diffuse-interface method, particularly the phase-field method, which is a type of interface-capturing method. 
The phase-field method is superior to other methods in terms of fluid mass conservation and its ability to implicitly handle dynamic topological changes of the interface. 
Consequently, it is widely employed in numerical simulations of two-phase flows~\cite{CHIU2011185,MIRJALILI2019221,MIRJALILI2020109006,MIRJALILI2021109918,JAIN2022111529,MIRJALILI2023111795,HWANG2024112972}. 
The modeling of surfactant transport using the phase-field method has been actively studied~\cite{KIM2006272,vanderSman2006,LIU20109166,ERIKTEIGEN2011375,GU2014416,Farsoiya2024,HAO2025114058,HAO2026}. 
In particular, the model proposed by Teigen et al.~\cite{ERIKTEIGEN2011375} is advantageous because of its simplicity and ability to conserve the surfactant mass. 
Further improvements to this model have been proposed, including methods to prevent unphysical negative concentrations caused by numerical errors~\cite{JAIN2024113277} and extensions that can accommodate realistic diffusion coefficients~\cite{HU2021106614,YAMASHITA2024113292}.

The purpose of this study is to propose approaches for further enhancing the accuracy of phase-field-based modeling of interfacial transport of insoluble surfactants. 
Specifically, we present the following two approaches:
\begin{itemize}
    \item \textbf{Adopting the formulation that avoids spatial derivatives of variables with sharp gradients.}
    We consider two types of transport models, both of which are generalized formulations of existing models.
    Although these formulations are mathematically equivalent at equilibrium, we demonstrate that one significantly outperforms the other in practical simulations because of reduced discretization error---a distinction that has not been emphasized in previous studies.
    \item \textbf{Decoupling the width of the delta function from the interface width.}
    We introduce a method that allows the width of the delta function to be specified independently of the interface width, enabling improved surfactant transport accuracy without sacrificing the interface-capturing quality or reducing the allowable time step.
    This strategy has not been systematically explored in previous diffuse-interface studies.
\end{itemize}
These two approaches are simple, effective, and practical in that they do not lead to significant increases in computational cost, implementation complexity, or degradation of interface-capturing accuracy.
They also preserve the discrete conservation of both fluid and surfactant mass.

Furthermore, we consider it problematic that most previous studies on interfacial surfactant transport have employed relatively simple benchmark tests. 
To address this issue, we propose a new benchmark test that poses substantial challenges in maintaining high accuracy and can serve as a rigorous standard for evaluating interfacial surfactant transport models.

The remainder of this paper is organized as follows.
In Section~\ref{sec:phasefield_and_surfactant}, we describe the phase-field method and two types of interfacial surfactant transport models. 
Section~\ref{sec:approaches} introduces the proposed approaches for improving their accuracy. 
Section~\ref{sec:overview_implementation} provides an overview of the implementation. 
Section~\ref{sec:numerical_tests} describes the numerical tests used to evaluate the effectiveness of the proposed approaches. 
Finally, Section~\ref{sec:conclusion} concludes the paper.

\section{Phase-field method and surfactant transport models}
\label{sec:phasefield_and_surfactant}

\subsection{Phase-field method}
\label{subsec:phasefield}

In this study, we use the phase-field method as an interface-capturing method for incompressible two-phase flows.
In particular, we adopt the accurate conservative diffuse-interface (ACDI) model~\cite{JAIN2022111529}, which is given by
\begin{equation}
    \frac{\partial \phi}{\partial t} + \nabla \cdot (\bm u \phi) = \nabla \cdot \gamma \left\{ \epsilon \nabla \phi - \frac{1}{4} \left[ 1 - \tanh^2\left(\frac{\psi_\mathrm{acdi}}{2 \epsilon}\right) \right] \frac{\nabla \psi_\mathrm{acdi}}{\left| \nabla \psi_\mathrm{acdi} \right|} \right\},
    \label{eq:acdi}
\end{equation}
where $\phi$ is the phase-field variable representing the volume fraction of one fluid, $\bm{u}$ is the velocity field, $\gamma$ is the velocity-scale parameter, and $\epsilon$ is the parameter controlling the interface width.
In this study, the interface width $W$ is defined as $W = 4 \epsilon$, following~\cite{YANG2021110113}.
To ensure $\phi$ remains numerically bounded between $0$ and $1$, the parameters must satisfy
\begin{equation}
    \gamma \geq \left| \bm{u} \right|_\mathrm{max} \quad \text{and} \quad \epsilon > 0.5 \Delta x, 
    \label{eq:boundedness_criterion}
\end{equation}
where $\left| \bm{u} \right|_\mathrm{max}$ is the maximum velocity magnitude and $\Delta x$ is the grid spacing~\cite{JAIN2022111529}.
In this study, we set $\gamma = 1.1 \left| \bm{u} \right|_\mathrm{max}$ and $\epsilon = 0.51 \Delta x$, resulting in an interface width of $W = 2.04 \Delta x$.
$\psi_\mathrm{acdi}$ is an auxiliary signed distance-like variable, given by
\begin{equation}
    \psi_\mathrm{acdi} = \epsilon \ln \left( \frac{\phi + \alpha_\psi}{1 - \phi + \alpha_\psi} \right),
    \label{eq:pf_to_sdf}
\end{equation}
where $\alpha_\psi = 10^{-100}$ is a small positive constant introduced to prevent $\psi_\mathrm{acdi}$ from diverging to $-\infty$ or $\infty$ as $\phi \to 0$ or $1$, respectively.
Equation~(\ref{eq:pf_to_sdf}) is derived from the equilibrium profile of the phase-field variable, given by
\begin{equation}
    \phi_\mathrm{eq} = \frac{1}{2} \left[ 1 + \tanh \left( \frac{\psi}{2 \epsilon} \right) \right],
\end{equation}
where $\psi$ denotes the signed distance from the interface.

We highlight several noteworthy features of the ACDI model~\cite{JAIN2022111529} in Eq.~(\ref{eq:acdi}):
\begin{itemize}
    \item Because the ACDI model is formulated as a conservative equation, solving it using the finite volume method ensures discrete conservation of fluid mass, which is essential for reliable simulations of two-phase flows. 
    \item The phase-field variable $\phi$ remains bounded between $0$ and $1$, provided that the parameters satisfy the criterion in Eq.~(\ref{eq:boundedness_criterion}) and the CFL condition, $\Delta t \leq \Delta x^2 / (2 N_\mathrm{dim} \gamma \epsilon)$, where $N_\mathrm{dim}$ is the number of spatial dimensions, and Eq.~(\ref{eq:acdi}) is appropriately discretized~\cite{JAIN2022111529}. 
    This boundedness property contributes to the robustness of the simulations.
    \item The ACDI model provides higher accuracy than the commonly used conservative diffuse-interface method proposed by Chiu and Lin~\cite{CHIU2011185}, as demonstrated in~\cite{JAIN2022111529}.
    \item The interface width $W (= 4\epsilon)$ should be set as small as possible, provided that the boundedness criterion in Eq.~(\ref{eq:boundedness_criterion}) is satisfied. 
    Increasing $\epsilon$ leads to a smaller allowable time step $\Delta t$ due to the CFL condition, which increases computational cost, and also degrades the accuracy of interface-capturing~\cite{JAIN2022111529}. 
    Therefore, we set $\epsilon = 0.51 \Delta x$, resulting in an interface width of approximately two grid cells ($W = 2.04 \Delta x$).
\end{itemize}

\subsection{Surfactant transport models}
\begin{figure}
    \centering
    \includegraphics[width=\linewidth]{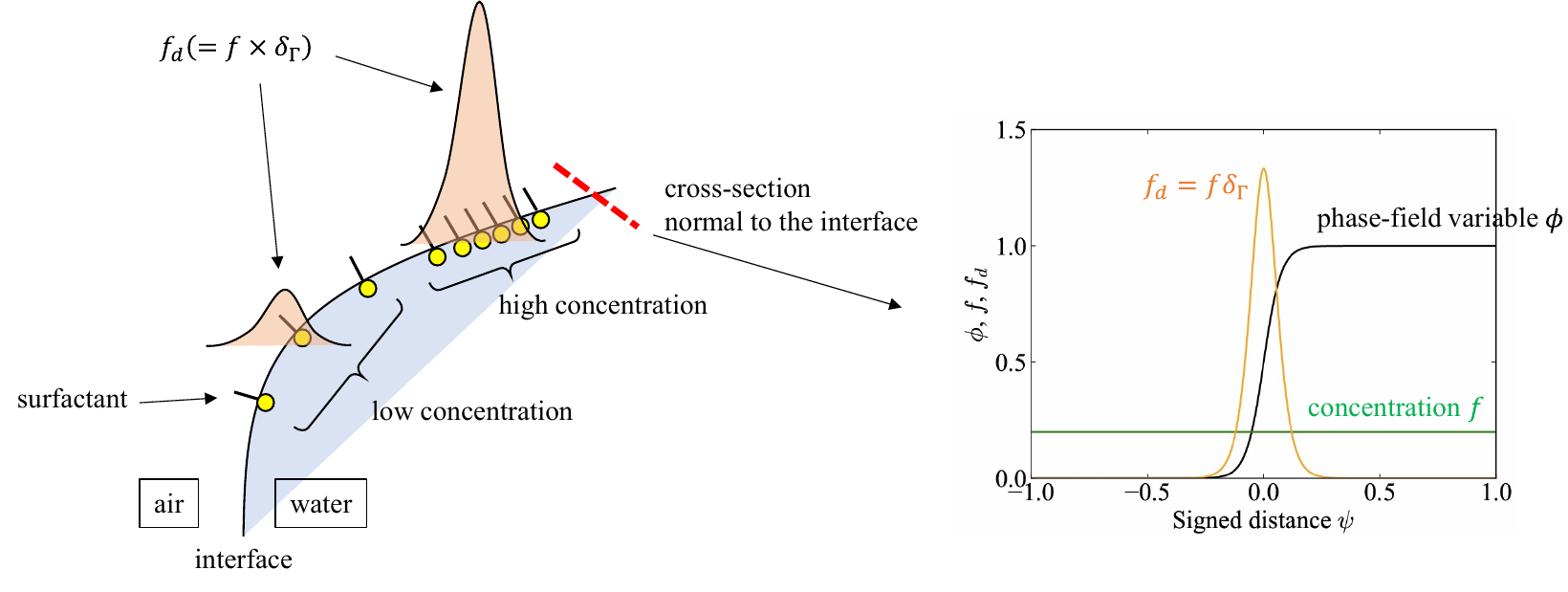}
    \caption{Schematic illustrating the representation of surfactant concentration using the diffuse-interface method.}
    \label{fig:schematic_profile_1D_concentration}
\end{figure}

Here, we describe interfacial surfactant transport models, that is, advection-diffusion equations of surfactant concentration on a deforming interface.
In this study, we compare two interfacial surfactant transport models.
The first model, proposed in~\cite{HU2021106614,YAMASHITA2024113292}, is given by
\begin{equation}
    \frac{\partial f_d}{\partial t} + \nabla \cdot (\bm u f_d) = \nabla \cdot D \left[ \nabla f_d - \frac{2 (0.5 - \phi) \bm n f_d}{\epsilon} \right] + \nabla \cdot \overline{D} \left[ \bm n \bm n^\top \nabla f_d - \frac{2 (0.5 - \phi) \bm n f_d}{\epsilon} \right],
    \label{eq:fd_type}
\end{equation}
where $f_d$ is the product of the surfactant concentration $f$ and the delta function $\delta_\Gamma$.
$f$ represents the interfacial surfactant concentration, that is, the amount of surfactant per unit area of the interface, while $f_d$ denotes the volumetric surfactant concentration, that is, the amount of surfactant per unit volume.  
At equilibrium, $f$ exhibits a flat profile in the direction normal to the interface~\cite{ERIKTEIGEN2011375,YAMASHITA2024113292}, whereas $f_d$ exhibits a sharp gradient in the normal direction, as illustrated in Fig.~\ref{fig:schematic_profile_1D_concentration}.
$\delta_\Gamma$ is the delta function, which takes large values within the interfacial region. 
This is often approximated by
\begin{equation}
    \delta_\Gamma = \frac{\phi (1 - \phi)}{\epsilon}.
    \label{eq:delta_wide}
\end{equation}
Note that in the original studies~\cite{HU2021106614,YAMASHITA2024113292}, the coefficients in front of the two $(0.5 - \phi)$ terms on the right-hand side of Eq.~(\ref{eq:fd_type}) are $4$ because they define the delta function as $\delta_\Gamma = 6 \phi^2 (1 - \phi)^2 / \epsilon$. 
In contrast, if $\delta_\Gamma$ is defined by Eq.~(\ref{eq:delta_wide}), the coefficients become $2$.
$D$ is the diffusion coefficient of the surfactant, and $\bm n$ is the interface normal vector. 
$\overline{D}$ is a freely chosen non-negative constant that controls how strongly the profile of $f_d$ is maintained, as detailed later in this section.
The second model, derived in~\cite{HU2021106614}, is given by
\begin{equation}
    \frac{\partial f_d}{\partial t} + \nabla \cdot (\bm u f_d) = \nabla \cdot \left( D \delta_\Gamma \nabla f \right) + \nabla \cdot \left( \overline{D} \delta_\Gamma \bm n \bm n^\top \nabla f \right).
    \label{eq:f_type}
\end{equation}
Both Eqs.~(\ref{eq:fd_type}) and (\ref{eq:f_type}) are conservative equations, and solving them using the finite volume method ensures the discrete conservation of the total surfactant mass.
The difference between Eqs.~(\ref{eq:fd_type}) and (\ref{eq:f_type}) lies on the right-hand side. 
The right-hand side of Eq.~(\ref{eq:fd_type}) includes $\nabla f_d$, whereas that of Eq.~(\ref{eq:f_type}) includes $\nabla f$.
In this study, Eq.~(\ref{eq:fd_type}) is referred to as the $f_d$-type model and Eq.~(\ref{eq:f_type}) as the $f$-type model, based on the forms of their right-hand sides.
As shown in \ref{sec:equivalence_f_and_fd} and in \cite{HU2021106614}, the $f_d$-type and $f$-type models are mathematically equivalent under the assumption that the phase-field variable $\phi$ maintains its equilibrium profile.
However, our numerical results reveal a notable difference in accuracy between the two models, with the $f$-type model exhibiting superior performance.  
A qualitative explanation for this observation is provided in the following section.

\begin{figure}
    \centering
    \includegraphics[width=\linewidth]{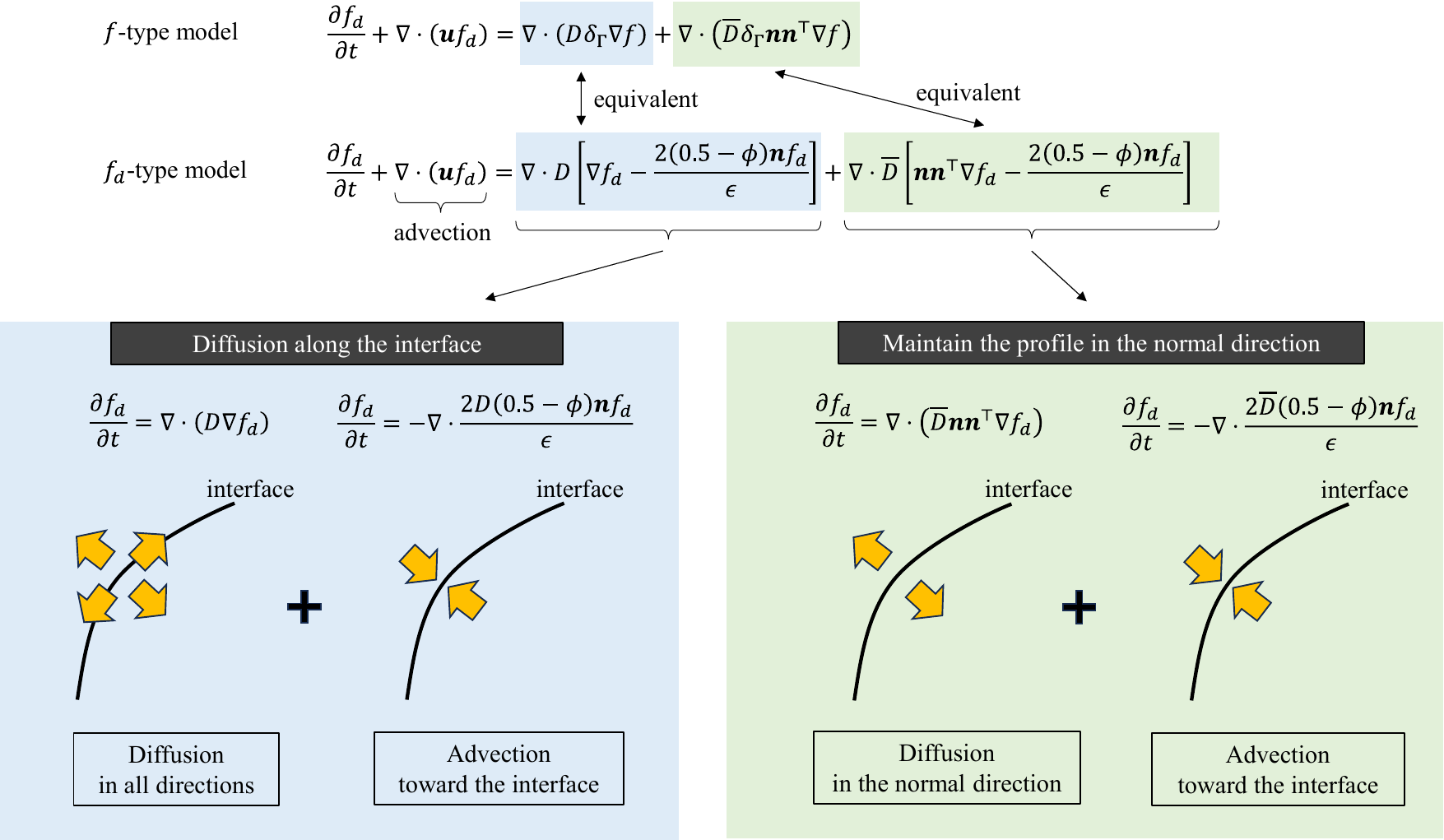}
    \caption{Schematic of the interfacial surfactant transport models compared in this study: the $f_d$-type model in Eq.~(\ref{eq:fd_type}) and the $f$-type model in Eq.~(\ref{eq:f_type}).}
    \label{fig:schematic_f_and_fd_type_models}
\end{figure}

Figure~\ref{fig:schematic_f_and_fd_type_models} illustrates the physical meaning of each term.
The left-hand side of the models represents the advection of the surfactant concentration by the velocity field $\bm{u}$.  
The first term on the right-hand side represents the surfactant diffusion along the interface, which consists of diffusion in all directions and an artificial sharpening term that advects the surfactant toward the interface.
The second term on the right-hand side maintains the profile of $f_d$ in the normal direction. 
It consists of the diffusion in the normal direction and an artificial sharpening term that advects the surfactant toward the interface.
The larger the coefficient $\overline{D}$, the more strongly the profile of $f_d$ is maintained in the normal direction.
We can set $\overline{D}$ sufficiently large, as long as the normal vector is accurately computed and the time step permitted by the CFL condition, $\Delta t \leq \Delta x^2 / \left(2 N_\mathrm{dim} \left(D + \overline{D}\right)\right)$, remains acceptable.

The $f_d$- and $f$-type models, originally derived by Hu~\cite{HU2021106614}, represent the generalized forms of several phase-field-based surfactant transport models:
\begin{itemize}
    \item Setting $\overline{D} = 0$ in the $f_d$-type model yields the model proposed by Jain~\cite{JAIN2024113277}.
    \item Setting $\overline{D} = \gamma \epsilon$ in the $f_d$-type model yields the model proposed by Yamashita et al.~\cite{YAMASHITA2024113292}.
    \item Setting $\overline{D} = 0$ in the $f$-type model yields the model proposed by Teigen et al.~\cite{ERIKTEIGEN2011375}.
\end{itemize}
In the following section, we present two approaches for improving the accuracy of both the $f_d$- and $f$-type models. 
These approaches are also effective for all the aforementioned models~\cite{ERIKTEIGEN2011375,HU2021106614,JAIN2024113277,YAMASHITA2024113292}.

\section{Approaches for enhancing the accuracy of surfactant transport}
\label{sec:approaches}

In this section, we propose two approaches to achieve highly accurate simulations of surfactant transport at the interface, based on the models introduced in Section~\ref{sec:phasefield_and_surfactant}.

\subsection{Approach~1: Using the $f$-type model instead of the $f_d$-type model}
\label{subsec:approach1}

The first approach is to use the $f$-type model in Eq.~(\ref{eq:f_type}), instead of the $f_d$-type model in Eq.~(\ref{eq:fd_type}).  
As demonstrated by the numerical tests in Section~\ref{sec:numerical_tests}, the $f$-type model yields more accurate results than the $f_d$-type model.  
Here, we provide a qualitative explanation of this observation.
The primary difference between the two models lies in their right-hand sides.  
The right-hand side of the $f_d$-type model includes the term $\nabla f_d$, which represents the spatial derivative of a variable with a sharp gradient in the direction normal to the interface.  
In contrast, the right-hand side of the $f$-type model includes $\nabla f$, which is the spatial derivative of a variable that has an approximately flat profile in the normal direction.  
The discretization error typically scales with the local gradient or wavenumber; thus, near the interface, evaluating $\nabla f_d$ is more prone to errors than $\nabla f$.
Consequently, the $f$-type model provides higher numerical accuracy than the $f_d$-type model.

\subsection{Approach~2: Decoupling the width of the delta function from the interface width}
\label{subsec:approach2}

\begin{figure}
    \centering
    \includegraphics[width=\linewidth]{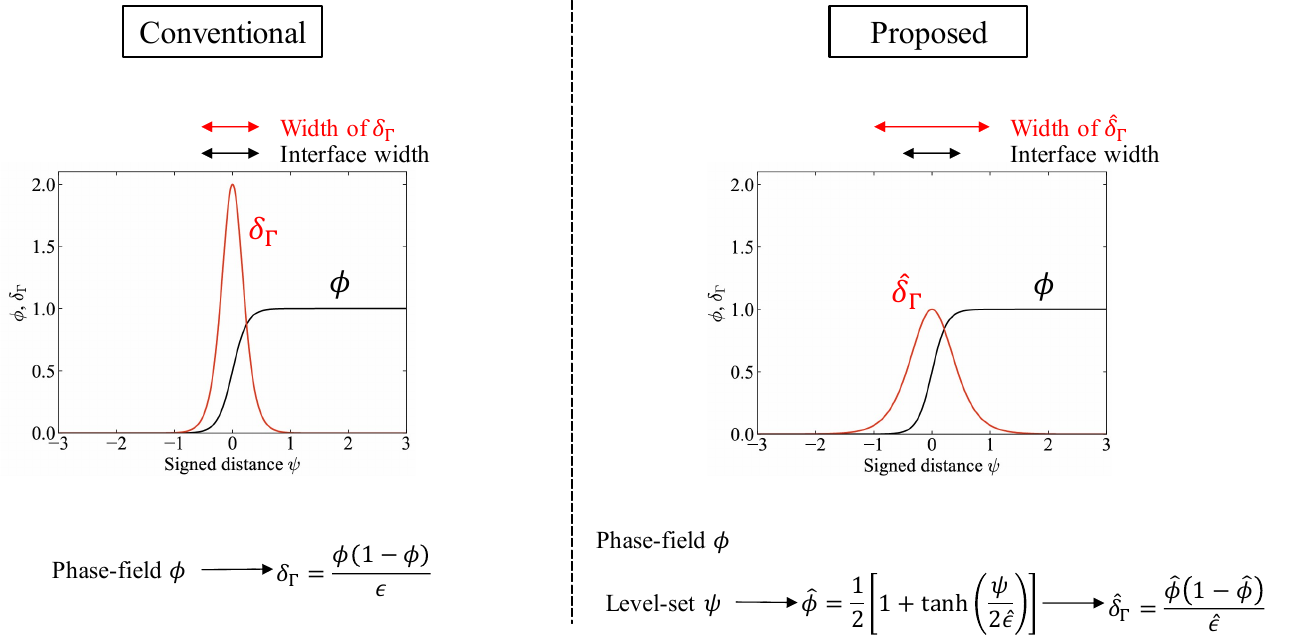}
    \caption{Schematic of the proposed Approach~2 in Section~\ref{subsec:approach2}. Conventionally, the delta function is computed directly from the phase-field variable $\phi$ using Eq.~(\ref{eq:delta_wide}), which ties the width of the delta function $\delta_\Gamma$ to the interface width. In the proposed approach, the delta function is computed from the level-set function $\psi$, allowing the width of the delta function to be adjusted independently of the interface width by controlling $\hat{\epsilon}$.}
    \label{fig:schematic_approach2}
\end{figure}

The second approach is to allow the width of the delta function $\delta_\Gamma$ to be adjusted independently of the interface width, as illustrated in Fig.~\ref{fig:schematic_approach2}.
As demonstrated by the numerical tests in Section~\ref{sec:numerical_tests}, the accuracy of surfactant transport is highly sensitive to the width of the delta function.  
In particular, the optimal width of the delta function is often larger than the preferred interface width.  
However, in conventional approaches, the delta function is computed directly from the phase-field variable using Eq.~(\ref{eq:delta_wide}), indicating that the delta function width is equal to the interface width.  
Consequently, increasing the width of the delta function requires an increase in the interface width, which, as discussed in Section~\ref{subsec:phasefield}, increases the computational cost and degrades the interface-capturing accuracy.  
Therefore, it is desirable to decouple the delta function width from the interface width, enabling their independent adjustment.

In this study, we employ the level-set method~\cite{SUSSMAN1994146} in conjunction with the ACDI method for interface capturing, enabling the width of the delta function to be specified independently of the interface width while conserving the fluid mass.
The level-set method represents the phases using a signed distance function $\psi$, referred to as the level-set function, where $\psi = 0$ denotes the interface and $\psi < 0$ and $\psi > 0$ correspond to $\phi < 0.5$ and $\phi > 0.5$, respectively.
In addition to updating the phase-field variable $\phi$ using Eq.~(\ref{eq:acdi}), the level-set function $\psi$ is advected at each time step using the following equation:
\begin{equation}
    \frac{\partial \psi}{\partial t} + \bm{u} \cdot \nabla \psi = 0.
    \label{eq:ls_advection}
\end{equation}
The level-set function $\psi$ should satisfy the following two conditions:
\begin{itemize}
    \item $\psi$ must retain the signed distance property; that is, $|\nabla \psi| = 1$.
    \item The interface locations defined by the level-set function and the phase-field variable must coincide; that is, $\psi = 0$ corresponds to $\phi = 0.5$.
\end{itemize}
To enforce these conditions, the reinitialization of $\psi$ is performed every 20 time steps as follows. 
First, $\psi_0$ is constructed by overwriting $\psi$ near the interface ($0.1 < \phi < 0.9$), using Eq.~(\ref{eq:pf_to_sdf}).  
The following reinitialization equation~\cite{SUSSMAN1994146} is then solved for 20 iterations with $\psi_0$ as the initial condition:
\begin{equation}
    \frac{\partial \psi}{\partial t_\mathrm{reinit}} + S(\psi_0) \left( \left| \nabla \psi \right| - 1 \right) = 0,
    \label{eq:reinit_ls}
\end{equation}
where $t_\mathrm{reinit}$ is the pseudo-time used only for this reinitialization process, and $S(\psi_0)$ denotes the sign function of $\psi_0$.
The reinitialization interval (every $20$ steps) and the number of iterations ($20$ iterations) are empirically determined so that the level-set function $\psi$ maintains the signed distance property without significant computational cost.

To enable a freely adjustable width for the delta function, we introduce a new parameter $\hat{\epsilon}$ that controls the width of the delta function independently of the interface width parameter $\epsilon$ used in the phase-field method.  
The delta function is computed using the level-set function $\psi$ and the parameter $\hat{\epsilon}$ as follows:
\begin{align}
    \hat{\phi} &= \frac{1}{2} \left[ 1 + \tanh \left( \frac{\psi}{2 \hat{\epsilon}} \right) \right], \label{eq:thick_pf} \\
    \hat{\delta_\Gamma} &= \left| \nabla \hat{\phi} \right| = \frac{\hat{\phi} (1 - \hat{\phi})}{\hat{\epsilon}}.
    \label{eq:thick_delta}
\end{align}
We define $\hat{W} = 4 \hat{\epsilon}$ as the width of the delta function, analogous to the interface width $W = 4 \epsilon$.  
By replacing $\epsilon$, $\phi$, and $\delta_\Gamma$ in the surfactant transport models [Eqs.~(\ref{eq:fd_type}) and (\ref{eq:f_type})] with $\hat{\epsilon}$, $\hat{\phi}$, and $\hat{\delta_\Gamma}$, respectively, we enable accurate simulations of surfactant transport.

The delta function width $\hat{W}$ should be set between approximately $3 \Delta x$ and $6 \Delta x$, while the interface width $W$ is fixed at $2.04 \Delta x$.
If the delta function is extremely narrow, the computational grid cannot resolve its sharp profile, resulting in large errors in surfactant transport simulations.
In contrast, if the delta function is extremely wide, it leads to degradation of interfacial surfactant diffusion, and the velocity gradients near the interface can distort the surfactant concentration profile.

The normal vector $\bm{n}$ used in Eqs.~(\ref{eq:fd_type}) and (\ref{eq:f_type}) is computed from the level-set function $\psi$ as
\begin{equation}
    \bm{n} = \frac{\nabla \psi}{\left| \nabla \psi \right|},
    \label{eq:normal_from_ls}
\end{equation}
which provides a more accurate representation of the interface normal than computing it from the phase-field variable, that is, $\nabla \phi / \left| \nabla \phi \right|$~\cite{JAIN2022111529}.

\subsection{Advantages of the proposed approaches}
The proposed approaches are simple and practical for enhancing the accuracy of surfactant transport simulations.
A key advantage is that they do not interfere with the phase-field method used for interface capturing.
Consequently, the accuracy of surfactant transport can be improved without increasing the interface width, thereby avoiding the degradation of the interface-capturing accuracy and the significant increase in computational cost associated with a reduced time step.
The additional computational cost of introducing the level-set method is negligible because in practical simulations of incompressible two-phase flows with surfactants, most of the computational time is spent solving the pressure Poisson equation~\cite{DODD2014416}.
Moreover, the proposed approaches preserve the discrete conservation of both fluid and surfactant mass.

\section{Overview of the implementation}
\label{sec:overview_implementation}

In this section, we present an overview of the implementation. 
A detailed implementation is provided in \ref{sec:detailed_implementation}. 
The governing equations are discretized on a uniform staggered Cartesian grid: all scalar quantities are defined at cell centers, whereas the velocity components are defined at cell faces. 
Time integration is performed using the third-order Runge-Kutta scheme~\cite{SHU1988439}, except for the level-set reinitialization.

The numerical procedure from time step $k$ to $k + 1$ is as follows:
\begin{enumerate}
    \item If $k \equiv 0 \pmod{20}$, reinitialize the level-set function $\psi^k$.
    \begin{itemize}
        \item Compute $\psi_0$ by overwriting $\psi^k$ near the interface $(0.1 < \phi^k < 0.9)$ using Eq.~(\ref{eq:pf_to_sdf}).
        \item Solve Eq.~(\ref{eq:reinit_ls}) for 20 iterations with $\psi_0$ as the initial condition, using the second-order Runge-Kutta scheme~\cite{SHU1988439}.
        A subcell fix~\cite{MIN20102764} is employed to prevent the interface from moving during this iteration.
        In cells whose centers are located within $0.01 \Delta x$ from the interface, we set $\psi^k = 0$, as described in~\cite{MIN2007300}.
        \item Update $\hat{\phi}^k$ using Eq.~(\ref{eq:thick_pf}).
    \end{itemize}
    \item Compute the phase-field variable $\phi^{k+1}$ using the ACDI method [Eq.~(\ref{eq:acdi})]. 
    The finite volume method is employed, as detailed in~\cite{JAIN2022111529}.
    \item Compute the level-set function $\psi^{k+1}$ by solving the advection equation [Eq.~(\ref{eq:ls_advection})]. 
    The finite difference method is used, where the spatial gradient is evaluated using the third-order weighted essentially non-oscillatory (WENO) scheme~\cite{Jiang_2000}.
    \item Compute $\hat{\phi}^{k+1}$ and $\hat{\delta_\Gamma}^{k+1}$ using the Eqs.~(\ref{eq:thick_pf}) and (\ref{eq:thick_delta}), respectively.
    \item Compute the surfactant concentrations $f_d^{k+1}$ and $f^{k+1}$.
    \begin{itemize}
        \item Compute $f_d^{k+1}$ using the surfactant transport model [Eq.~(\ref{eq:fd_type}) or Eq.~(\ref{eq:f_type})], replacing $\epsilon$, $\phi$, and $\delta_\Gamma$ with $\hat{\epsilon}$, $\hat{\phi}$, and $\hat{\delta_\Gamma}$, respectively.
        The finite volume method is used for discretization, where the flux at cell faces is computed using linear interpolation and the second-order central difference scheme.
        \item Obtain $f^{k+1}$ by dividing $f_d^{k+1}$ by $\hat{\delta_\Gamma}^{k+1}$.
    \end{itemize}
    \item Compute the normal vector $\bm{n}^{k+1}$ using Eq.~(\ref{eq:normal_from_ls}).
    \item Update $\bm{u}^{k+1}$ by applying a prescribed velocity field or solving the fluid equations.
    \item Update the parameter $\gamma^{k+1} = 1.1 \left| \bm{u}^{k+1} \right|_\mathrm{max}$ in the ACDI model [Eq.~(\ref{eq:acdi})].
    \item Advance time by setting $t = t + \Delta t$, and return to Step 1 with $k = k + 1$.
\end{enumerate}

\section{Numerical tests}
\label{sec:numerical_tests}

In this section, we present numerical tests to verify that the approaches described in Section~\ref{sec:approaches} enhance the accuracy of interfacial surfactant transport.  
Finally, we introduce a challenging problem that is difficult to solve accurately and thus serves as a valuable benchmark for evaluating and comparing the performance of various methods proposed in existing studies.

\subsection{Surfactant diffusion in 2D uniform flow}
\label{subsec:diffusion_in_uniform_flow}

\begin{figure}
    \centering
    \includegraphics[width=0.8\linewidth]{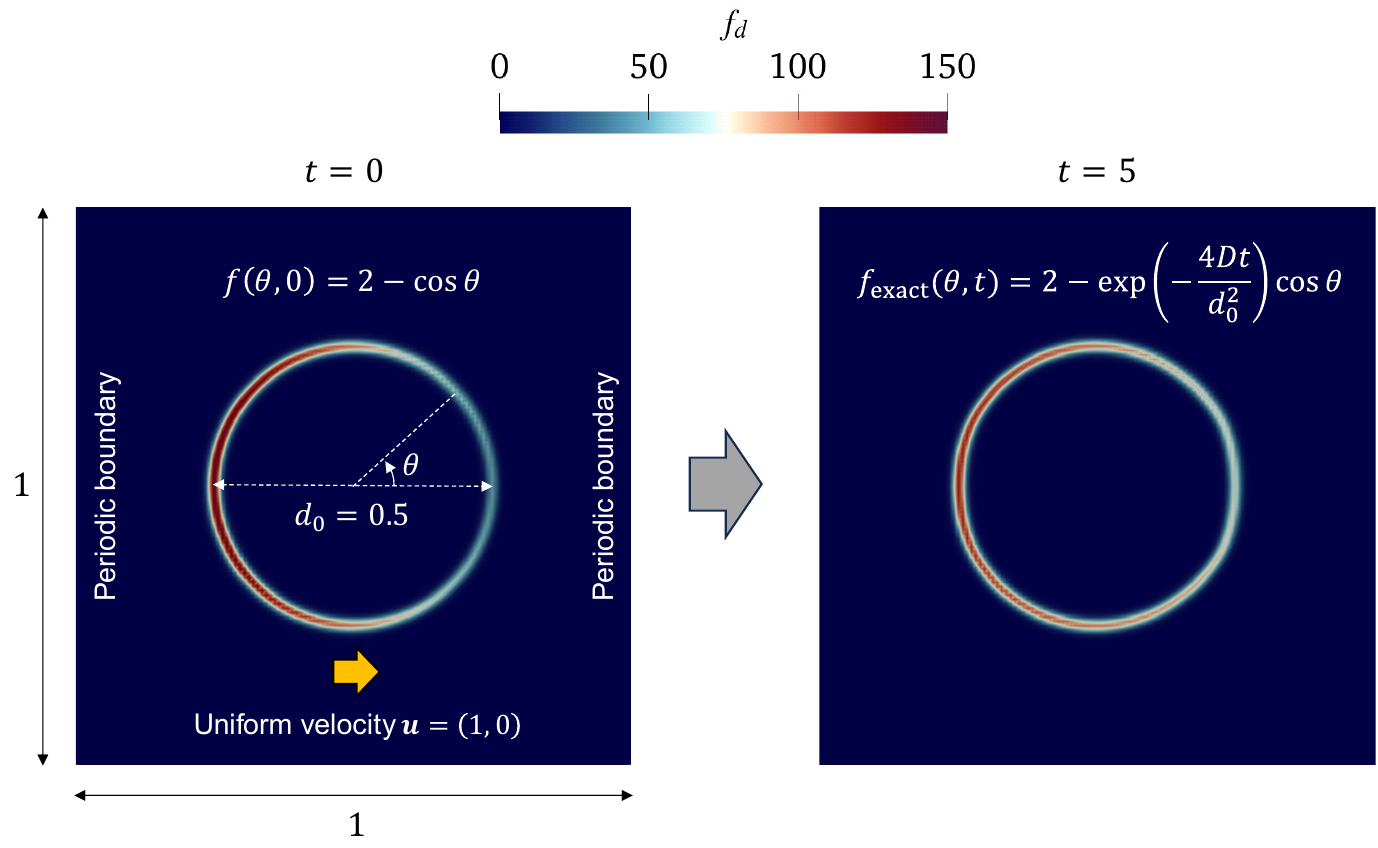}
    \caption{Schematic of surfactant diffusion on a circular interface advected by a uniform velocity field, as described in Section~\ref{subsec:diffusion_in_uniform_flow}.}
    \label{fig:schematic_diffusion_in_uniform_flow}
\end{figure}

In this test case, the surfactant diffuses along a circular interface advected by a uniform velocity field.  
As shown in Fig.~\ref{fig:schematic_diffusion_in_uniform_flow}, a circular interface with a diameter of $d_0 = 0.5$ is initially centered within a square domain of size $1 \times 1$.  
Periodic boundary conditions are imposed on the left and right boundaries of the domain.  
The surfactant is initially distributed along the interface at a concentration given by  
\begin{equation}
    f(\theta, t = 0) = 2 - \cos \theta.
\end{equation}
The interface is advected by the uniform velocity field $\bm{u} = (1, 0)$, while the surfactant diffuses along the interface.
The exact solution for the surfactant concentration at time $t$ is given by  
\begin{equation}
    f_\mathrm{exact}(\theta, t) = 2 - \exp \left(- \frac{4 D t}{d_0^2}\right) \cos \theta.
\end{equation}
We perform this test using both the $f_d$-type model [Eq.~(\ref{eq:fd_type})] and $f$-type model [Eq.~(\ref{eq:f_type})], varying the delta function width $\hat{W}$, and using two diffusion coefficients ($D = 10^{-2}$ and $10^{-9}$) to answer the following questions:
\begin{itemize}
\item Are the proposed approaches effective across a wide range of advection-to-diffusion ratios?
\item How does the delta function width $\hat{W}$ affect the numerical accuracy?
\end{itemize}  
The parameter $\overline{D}$ in the surfactant transport models [Eqs.~(\ref{eq:fd_type}) and (\ref{eq:f_type})] is set to $\overline{D} = 10^{-2}$.
A fixed time step of $\Delta t = 10^{-4}$ is used in all the cases.

\subsubsection*{Effectiveness of Approach~1 in Section~\ref{subsec:approach1}}
\begin{figure}
    \centering
    \includegraphics[width=\linewidth]{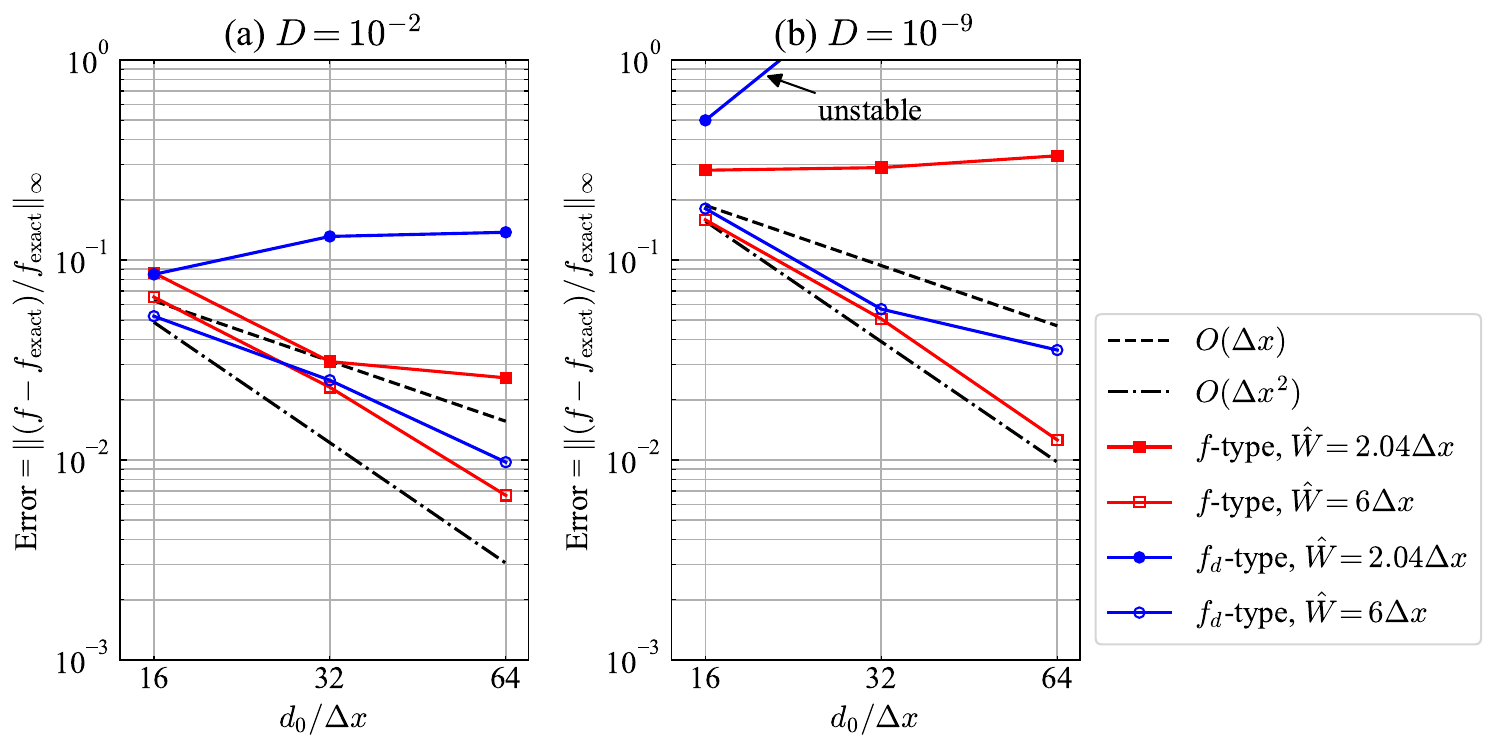}
    \caption{Error in surfactant concentration at $t = 5$ for the case of surfactant diffusion in uniform flow, as described in Section~\ref{subsec:diffusion_in_uniform_flow}. Two different diffusion coefficients are examined: (a) $D = 10^{-2}$ and (b) $D = 10^{-9}$. The test is performed using three different grid resolutions: $32^2$, $64^2$, and $128^2$, corresponding to $16$, $32$, and $64$ grid points per diameter $d_0$, respectively. $\hat{W}$ denotes the width of the delta function.}
    \label{fig:diffusion_in_uniform_flow_convergence}
\end{figure}

The results of the grid convergence study for the error in the surfactant concentration at $t = 5$ are shown in Fig.~\ref{fig:diffusion_in_uniform_flow_convergence}.
Three different grid resolutions, $32^2$, $64^2$, and $128^2$, are used in the simulations, corresponding to $16$, $32$, and $64$ grid points per diameter $d_0$, respectively. 
The error is calculated as $\left\| (f - f_\mathrm{exact}) / f_\mathrm{exact} \right\|_\infty$, where the concentration $f$ is extracted at the interface using ParaView's Contour filter~\cite{AHRENS2005717}.
For both a large diffusion coefficient ($D = 10^{-2}$) and a small diffusion coefficient ($D = 10^{-9}$), the $f$-type model [Eq.~(\ref{eq:f_type})] outperforms the $f_d$-type model [Eq.~(\ref{eq:fd_type})] when using the same width for the delta function: the $f$-type model consistently yields smaller errors in almost all cases. 
Furthermore, for $D = 10^{-9}$ and the width of the delta function $\hat{W} = 2.04 \Delta x$, the $f_d$-type model becomes unstable at high resolutions, whereas the $f$-type model remains stable. 
These findings indicate that the $f$-type model is superior to the $f_d$-type model in terms of both accuracy and stability; thus, the proposed Approach~1 is effective across a wide range of advection-to-diffusion ratios.

\subsubsection*{Effectiveness of Approach~2 in Section~\ref{subsec:approach2}}
Fig.~\ref{fig:diffusion_in_uniform_flow_convergence} demonstrates that, in all cases, allowing $\hat{W}$ to be freely chosen can significantly improve the accuracy compared to fixing $\hat{W}$ to the interface width $W\ (= 2.04 \Delta x)$.
In the most accurate configuration, where the $f$-type model is employed with $\hat{W} = 6 \Delta x$, nearly second-order accuracy is achieved for both diffusion coefficients.
Additionally, for $D = 10^{-9}$, the $f_d$-type model with $\hat{W} = 2.04 \Delta x$ becomes unstable at high resolution, whereas $\hat{W} = 6 \Delta x$ remains stable, and the error converges. 
These results indicate that allowing $\hat{W}$ to be freely chosen can significantly improve the accuracy and robustness of the simulations; thus, the proposed Approach~2 is effective across a wide range of advection-to-diffusion ratios.

\subsubsection*{Effect of the delta function width on numerical accuracy}
\begin{figure}
    \centering
    \includegraphics[width=\linewidth]{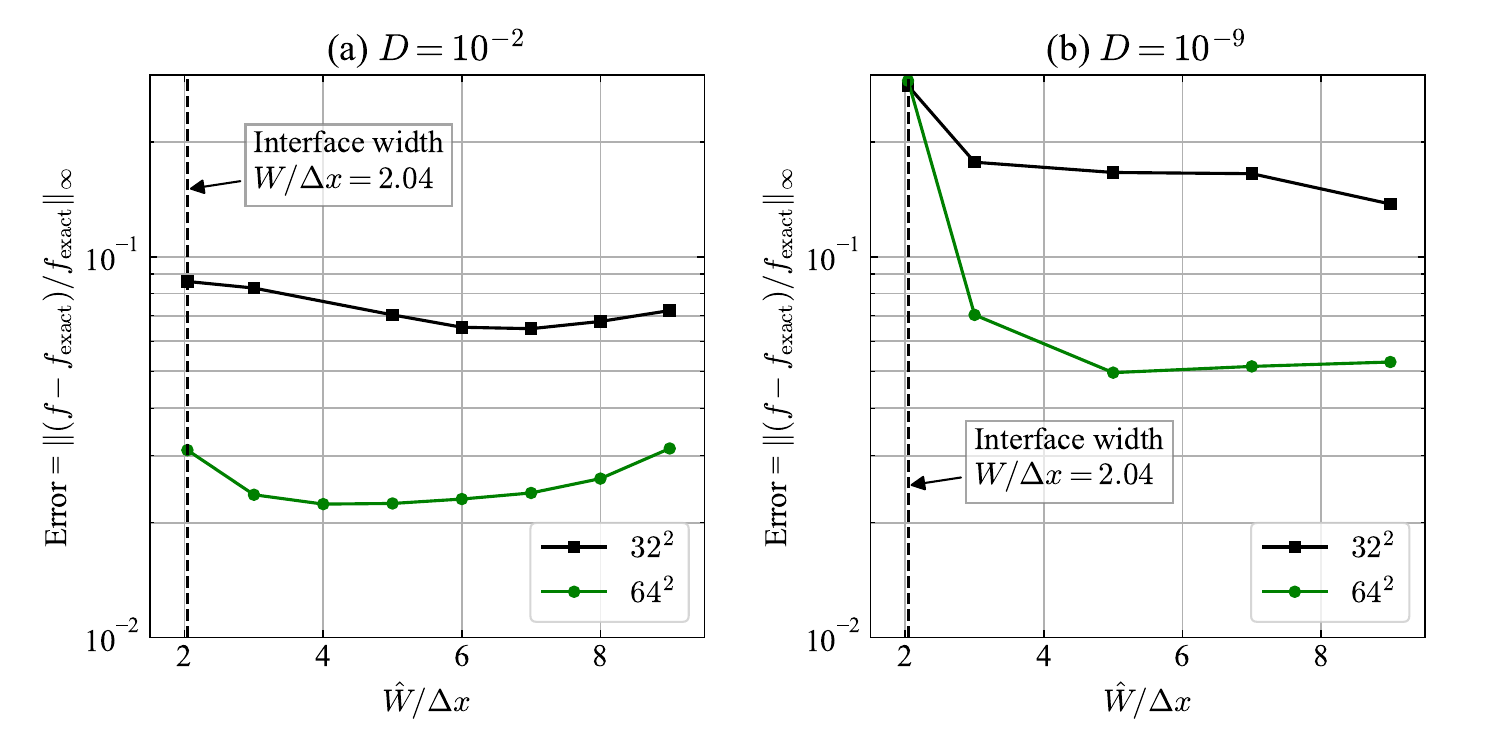}
    \caption{Errors for various widths of the delta function $\hat{W}$ when using an $f$-type model to simulate surfactant diffusion in a uniform flow, as described in Section~\ref{subsec:diffusion_in_uniform_flow}. Panel (a) shows results for a diffusion coefficient $D = 10^{-2}$, and panel (b) for $D = 10^{-9}$. Simulations were conducted with two different grid resolutions: $32^2$ and $64^2$.}
    \label{fig:diffusion_in_uniform_flow_error_vs_width}
\end{figure}

Figure~\ref{fig:diffusion_in_uniform_flow_error_vs_width} shows the error in the surfactant concentration for various widths of the delta function when the $f$-type model is used to simulate surfactant diffusion in a uniform flow. 
As shown in Fig.~\ref{fig:diffusion_in_uniform_flow_error_vs_width}(a), for a large diffusion coefficient ($D = 10^{-2}$), an optimal width exists for the delta function that minimizes error.
This behavior reflects a trade-off between two competing sources of error: one arising from an excessively narrow delta function and the other from an overly wide one.  
If the delta function is extremely narrow, the computational grid cannot adequately resolve its sharp profile, resulting in significant errors in both the advection and diffusion terms.  
In contrast, if the delta function is extremely wide, a substantial portion of the surfactant concentration $f_d$ is distributed away from the interface, thereby degrading the accuracy of the surfactant diffusion along the interface.  
For a small diffusion coefficient, such as $D = 10^{-9}$, the error caused by the degraded diffusion is relatively minor compared to the advection error, even when the delta function has a large width, as shown in Fig.~\ref{fig:diffusion_in_uniform_flow_error_vs_width}(b).  
Therefore, increasing the width of the delta function does not significantly amplify the error when the diffusion coefficient is small.

\subsubsection*{Conservation of surfactant mass}
\begin{figure}
    \centering
    \includegraphics[width=0.6\linewidth]{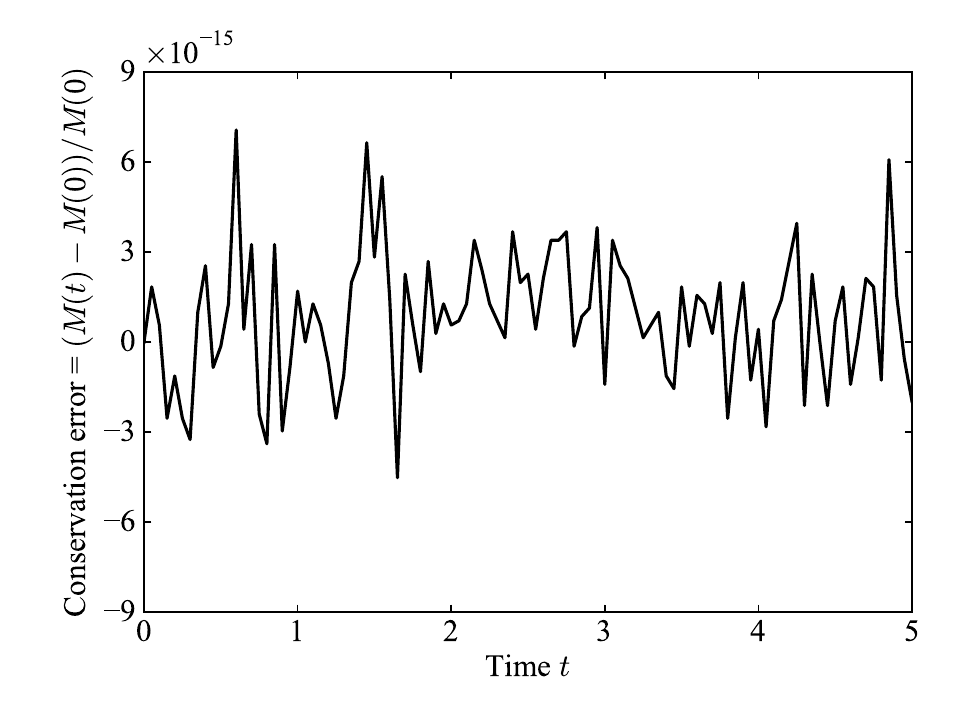}
    \caption{
    Time evolution of the conservation error of surfactant mass for the diffusion test in a 2D uniform flow (see Section~\ref{subsec:diffusion_in_uniform_flow}). The conservation error is defined in Eq.~(\ref{eq:conservation_error}). The diffusion coefficient is $D = 10^{-2}$. The grid resolution is $64^2$. 
    The $f$-type model with $\hat{W} = 6 \Delta x$ is used.}
    \label{fig:diffusion_in_uniformflow_conservation_error}
\end{figure}

We demonstrate the conservation of surfactant mass for $D = 10^{-2}$ on a $64^2$ grid.
Figure~\ref{fig:diffusion_in_uniformflow_conservation_error} shows the conservation error of surfactant mass for the $f$-type model with $\hat{W} = 6 \Delta x$.
The conservation error is defined as
\begin{equation}
    \mathrm{Error} = \frac{M(t) - M(0)}{M(0)}, \label{eq:conservation_error}
\end{equation}
where $M(t)$ is the total surfactant mass at time $t$, given by
\begin{align}
    M(t) = \sum_{i,j} (f_d)_{i, j} \Delta x \Delta y.
\end{align}
The error is on the order of $10^{-15}$, indicating that the proposed approaches are discretely conservative.

\subsection{Surfactant transport in 2D vortical flow}
\label{subsec:single_vortex}

\begin{figure}
    \centering
    \includegraphics[width=\linewidth]{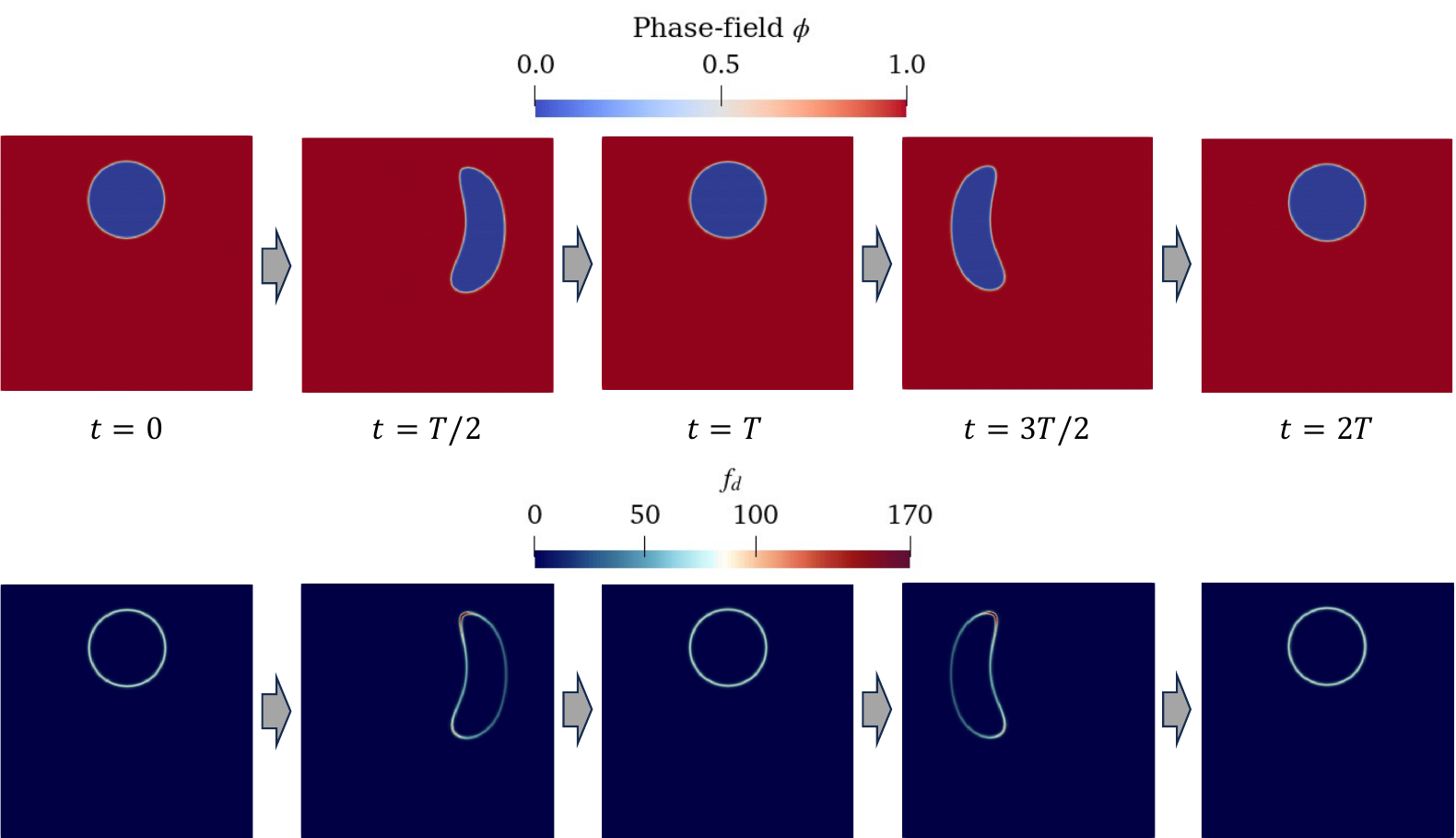}
    \caption{Schematic of surfactant transport in a 2D vortical flow, as described in Section~\ref{subsec:single_vortex}. A bubble is deformed by the periodic vortical velocity field given in Eq.~(\ref{eq:vortex2D_velocity}) with $T = 1$. The bubble shape and the surfactant concentration ideally return to their initial states at $t = T$ and $t = 2T$. The top row shows the phase-field variable $\phi$, while the bottom row shows the surfactant concentration $f_d$.}
    \label{fig:schematic_single_vortex}
\end{figure}

In this test, we simulate surfactant transport along an interface deformed by a 2D vortical flow. 
This vortical flow scenario is commonly used to evaluate the accuracy of interface-capturing methods~\cite{RIDER1998112,MIRJALILI2019221,JAIN2022111529}, and it is employed in this study to assess the performance of the surfactant transport models.
As shown in Fig.~\ref{fig:schematic_single_vortex}, at time $t = 0$, a bubble with diameter $d_0 = 0.3$ centered at $(0.5, 0.75)$ is placed within the domain $[0, 1] \times [0, 1]$. 
The surfactant is initially distributed along the bubble interface at a uniform concentration $f(\theta, t = 0) = 1$.
The bubble and surfactant are advected by a vortical velocity field $\bm u = (u, v)$~\cite{JAIN2022111529}:
\begin{subequations}
    \begin{align}
        u(x, y, t) &= - \sin^2(\pi x) \sin(2\pi y) \cos \left( \frac{\pi t}{T} \right), \\
        v(x, y, t) &= \sin(2 \pi x) \sin^2 (\pi y) \cos \left( \frac{\pi t}{T} \right),
    \end{align}
    \label{eq:vortex2D_velocity}
\end{subequations}
where $T = 1$ in this test case.
Because the velocity field is periodically inverted, the interface ideally returns to its original shape at times $t = T$ and $t = 2T$. 
In this test, the surfactant diffusion coefficient is set to $D = 0$; therefore, the surfactant concentration is transported purely by advection and should ideally return to its initial state at $t = T$ and $t = 2T$. 
The simulation is performed using a fixed time step of $\Delta t = 10^{-4}$.
The parameter $\overline{D}$ in the surfactant transport models [Eqs.~(\ref{eq:fd_type}) and (\ref{eq:f_type})] is set to $\overline{D} = 10^{-2}$.

\subsubsection*{Effectiveness of Approach~1 in Section~\ref{subsec:approach1}}
\begin{figure}
    \centering
    \includegraphics[width=0.8\linewidth]{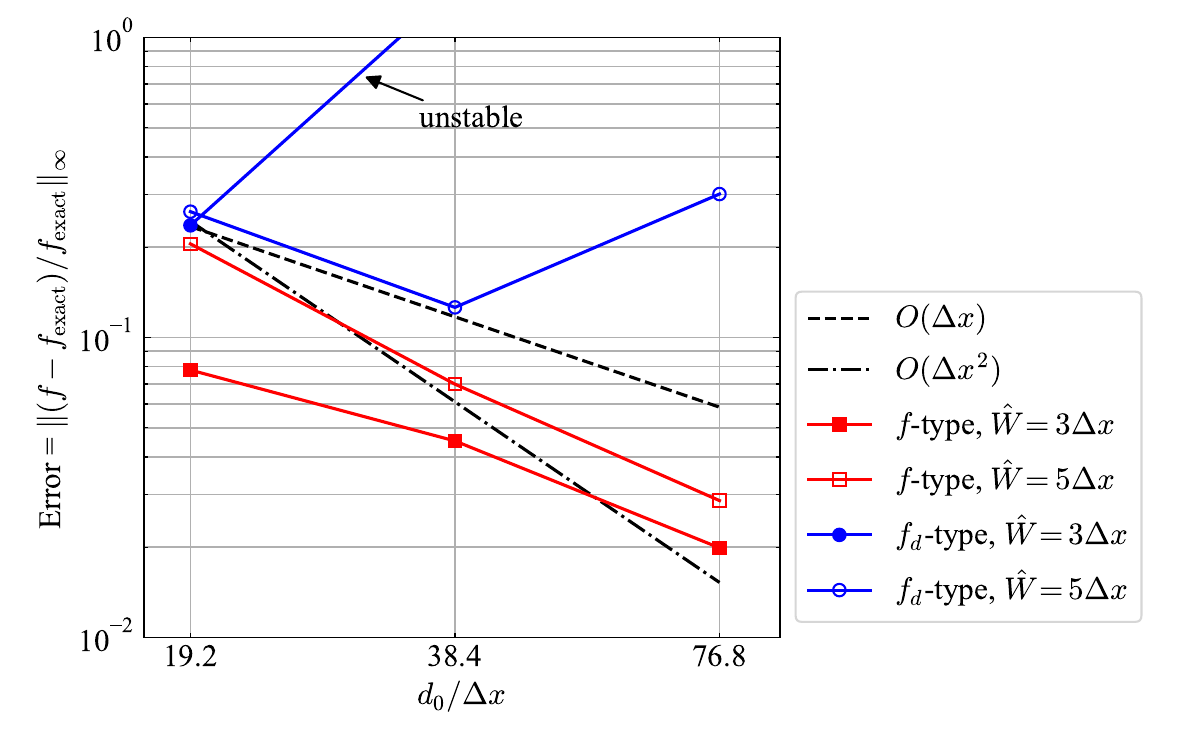}
    \caption{Errors in surfactant concentration at $t = 2T$ for the 2D vortical flow test described in Section~\ref{subsec:single_vortex}. Simulations are performed using different widths $\hat{W}$ of the delta function for both the $f_d$-type and $f$-type models. The grid resolutions are $64^2$, $128^2$, and $256^2$, corresponding to $19.2$, $38.4$, and $76.8$ grid points per bubble diameter $d_0$, respectively.}
    \label{fig:single_vortex_convergence}
\end{figure}

Figure~\ref{fig:single_vortex_convergence} shows the results of the grid convergence study for the two models (the $f_d$-type model [Eq.~(\ref{eq:fd_type})] and $f$-type model [Eq.~(\ref{eq:f_type})]) and two different delta function widths ($\hat{W} = 3\Delta x$ and $\hat{W} = 5\Delta x$).
The grid resolutions are $64^2$, $128^2$, and $256^2$, corresponding to $19.2$, $38.4$, and $76.8$ grid points per bubble diameter $d_0$, respectively.
At all resolutions, the $f$-type model exhibits smaller errors than the $f_d$-type model. 
Furthermore, for $\hat{W} = 3\Delta x$, the $f_d$-type model becomes unstable at higher resolutions, whereas the $f$-type model remains stable. 
These results indicate that the $f$-type model is superior to the $f_d$-type model in terms of both accuracy and stability; thus, the proposed Approach~1 is effective even in the presence of interface deformation.

\subsubsection*{Effectiveness of Approach~2 in Section~\ref{subsec:approach2}}
As shown in Fig.~\ref{fig:single_vortex_convergence}, the $f_d$-type model becomes unstable at high resolution when a narrow delta function ($\hat{W} = 3 \Delta x$) is used, whereas it remains stable with a wider delta function ($\hat{W} = 5 \Delta x$). 
Thus, the proposed Approach~2 contributes to the stability of the simulation.

\subsubsection*{Effect of the delta function width on numerical accuracy}
\begin{figure}
    \centering
    \includegraphics[width=0.7\linewidth]{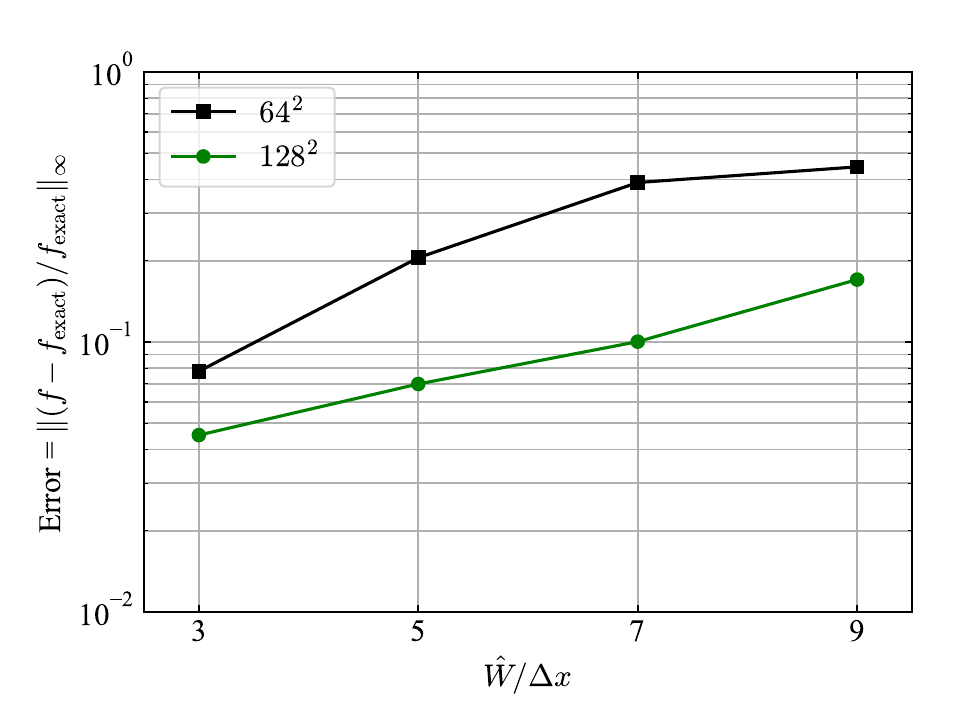}
    \caption{Errors in surfactant concentration for various widths of the delta function in the 2D vortical flow test described in Section~\ref{subsec:single_vortex}. Results are shown for the $f$-type model at two grid resolutions: $64^2$ and $128^2$.}
    \label{fig:single_vortex_error_vs_width}
\end{figure}

Figure~\ref{fig:single_vortex_error_vs_width} shows the relationship between the width of the delta function and the error in the surfactant concentration at $t = 2T$ when the $f$-type model is used. 
The errors are plotted for two grid resolutions: $64^2$ and $128^2$. 
For both resolutions, the error increases as the delta function width increases.  
This behavior arises because a wide delta function leads to the distortion of the surfactant concentration profile due to the velocity gradients near the interface. 
In contrast, in the uniform flow case described in Section~\ref{subsec:diffusion_in_uniform_flow}, where no velocity gradient is present, increasing the width $\hat{W}$ does not significantly increase the error when the diffusion coefficient is small.  
However, due to the strong velocity gradients in the present test case, increasing the delta function width leads to larger errors, even in the absence of diffusion.

\subsection{Surfactant transport in 3D vortical flow}
\label{subsec:vortex3D}

\begin{figure}
    \centering
    \includegraphics[width=\linewidth]{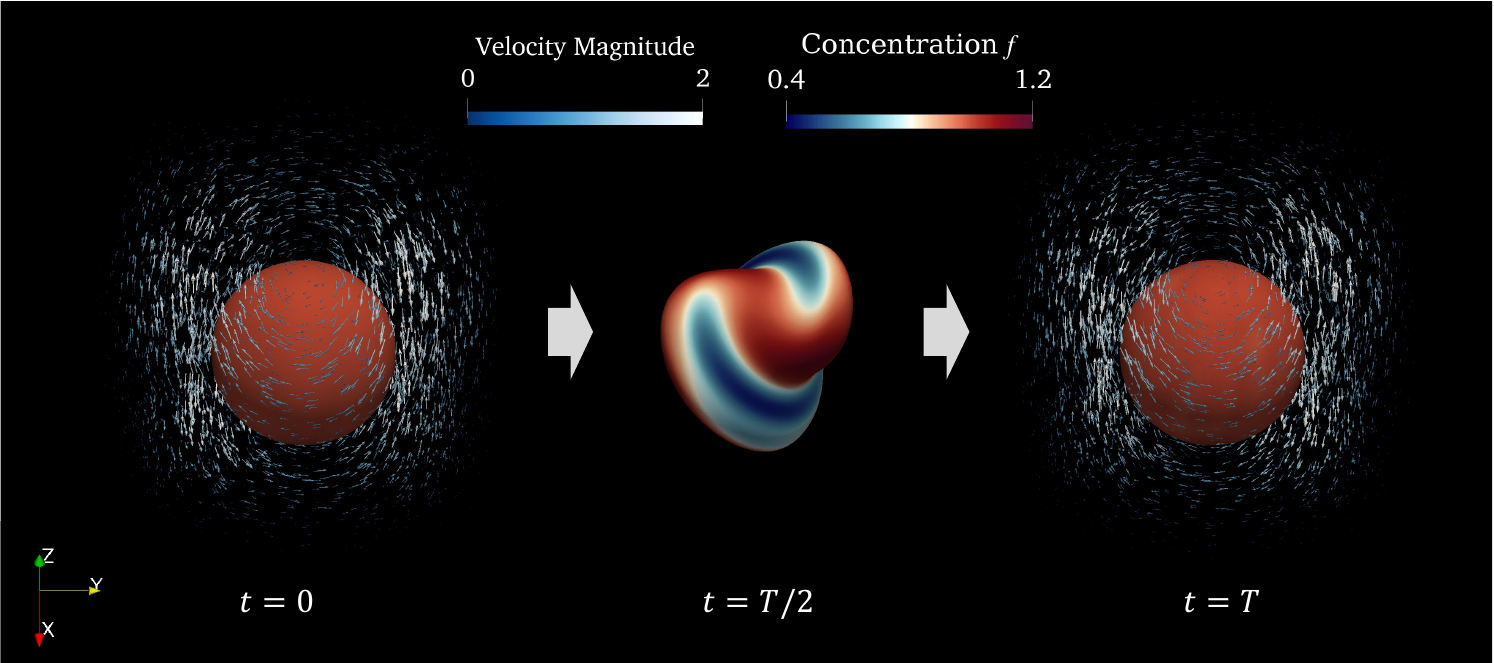}
    \caption{Schematic of surfactant transport in a 3D vortical flow, as described in Section~\ref{subsec:vortex3D}. A drop is deformed by the periodic vortical velocity field given in Eq.~(\ref{eq:vortex3D_velocity}). The drop shape and the surfactant concentration ideally return to their initial states at $t = T$. The arrows indicate the velocity field, and the color on the drop interface represents the surfactant concentration $f$.}
    \label{fig:schematic_vortex3D}    
\end{figure}

To demonstrate that the conclusions from the 2D test cases remain valid in 3D, we simulate surfactant transport on a drop interface deformed by a 3D vortical flow, as shown in Fig.~\ref{fig:schematic_vortex3D}.
Initially, a drop with diameter $d_0 = 0.5$ is placed at $(0.5, 0.5, 0.4)$ within the computational domain $[0, 1] \times [0, 1] \times [0, 1]$.
The surfactant is initially distributed uniformly on the drop interface, such that $f(x, y, z, t = 0) = 1$.
Surfactant diffusion is neglected, i.e., $D = 0$.
The drop is deformed by a prescribed periodic velocity field~\cite{Leveque_1996}:
\begin{subequations}
    \begin{align}
        u(x, y, z, t) &= 2 \sin^2(\pi x) \sin(2 \pi y) \sin(2 \pi z) \cos\left( \frac{\pi t}{T} \right), \\
        v(x, y, z, t) &= - \sin(2 \pi x) \sin^2(\pi y) \sin(2 \pi z) \cos\left( \frac{\pi t}{T} \right), \\
        w(x, y, z, t) &= - \sin(2 \pi x) \sin(2 \pi y) \sin^2(\pi z) \cos\left( \frac{\pi t}{T} \right),
    \end{align}
    \label{eq:vortex3D_velocity}
\end{subequations}
where $T = 1$ in this test.
Similar to the previous 2D vortical flow test, the velocity field inverts at $t = T/2$.
Consequently, at $t = T$, the drop shape and the surfactant concentration $f$ ideally return to their initial states (see Fig.~\ref{fig:schematic_vortex3D}).
The simulations use $\overline{D} = 10^{-2}$ and a time step of $\Delta t = 10^{-4}$.

\subsubsection*{Effectiveness of Approach~1 in Section~\ref{subsec:approach1}}

\begin{figure}
    \centering
    \includegraphics[width=0.7\linewidth]{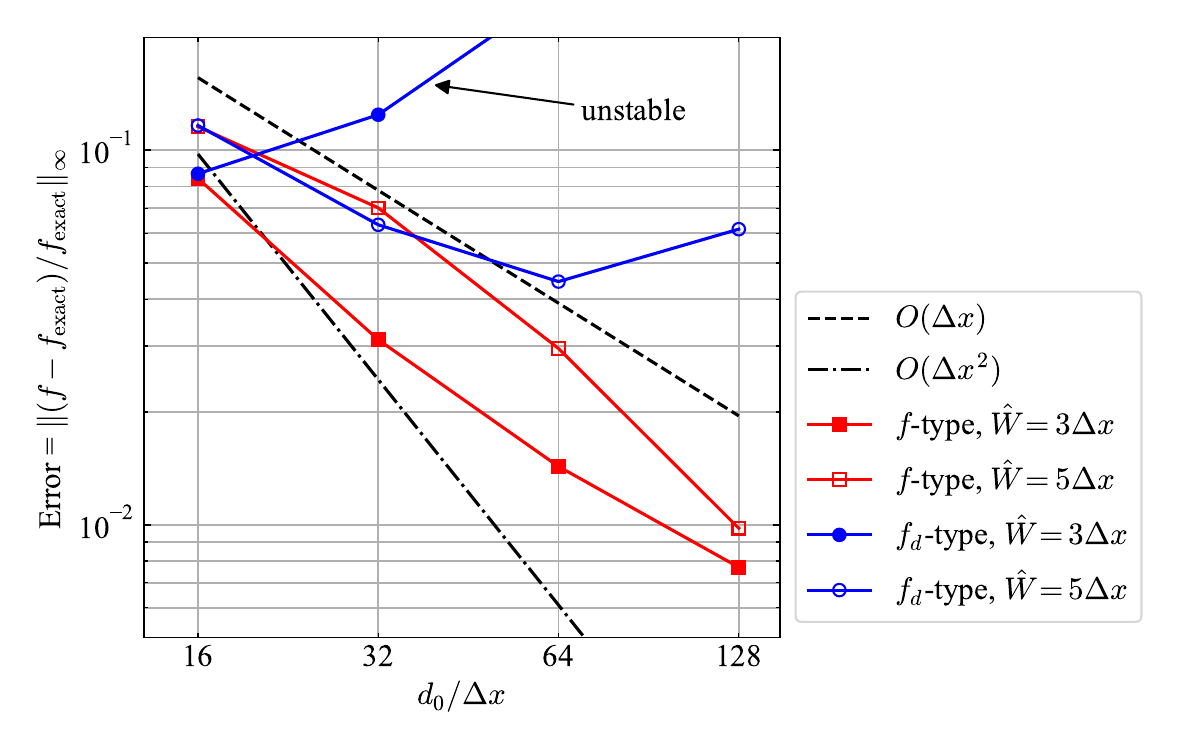}
    \caption{Errors in the surfactant concentration at $t = T$ for the 3D vortical flow test described in Section~\ref{subsec:vortex3D}. Simulations are performed using different delta function widths $\hat{W}$ for both the $f_d$-type and $f$-type models. The grid resolutions are $32^3$, $64^3$, $128^3$, and $256^3$, corresponding to $16$, $32$, $64$, and $128$ grid points per drop diameter $d_0$, respectively.}
    \label{fig:vortex3D_convergence}
\end{figure}

Figure~\ref{fig:vortex3D_convergence} presents the results of the grid convergence study, comparing errors in the surfactant concentration $f$ between two models (the $f_d$-type model~[Eq.~(\ref{eq:fd_type})] and the $f$-type model~[Eq.~(\ref{eq:f_type})]) and two delta function widths ($\hat{W} = 3 \Delta x$ and $\hat{W} = 5 \Delta x$).
The grid resolutions are $32^3$, $64^3$, $128^3$, and $256^3$, corresponding to $16$, $32$, $64$, and $128$ grid points per drop diameter $d_0$, respectively.
For both delta function widths, the $f$-type model yields smaller errors and better convergence rates than the $f_d$-type model.
In particular, for $\hat{W} = 3 \Delta x$, the $f_d$-type model becomes unstable, whereas the $f$-type model remains stable and the errors converge as the resolution increases.
These observations indicate that the $f$-type model is superior to the $f_d$-type model in both accuracy and stability; thus, the proposed Approach~1 remains effective in 3D as well as in 2D.

\subsection*{Effectiveness of Approach~2 in Section~\ref{subsec:approach2}}

As shown in Fig.~\ref{fig:vortex3D_convergence}, the $f_d$-type model becomes unstable at high resolution when using a narrow delta function ($\hat{W} = 3 \Delta x$), but remains stable with a wider delta function ($\hat{W} = 5 \Delta x$). This observation suggests that the proposed Approach~2 enhances the stability of this 3D vortical flow test, consistent with its performance in the previous 2D case.

\subsection*{Effect of the delta function width on numerical accuracy}

\begin{figure}
    \centering
    \includegraphics[width=0.7\linewidth]{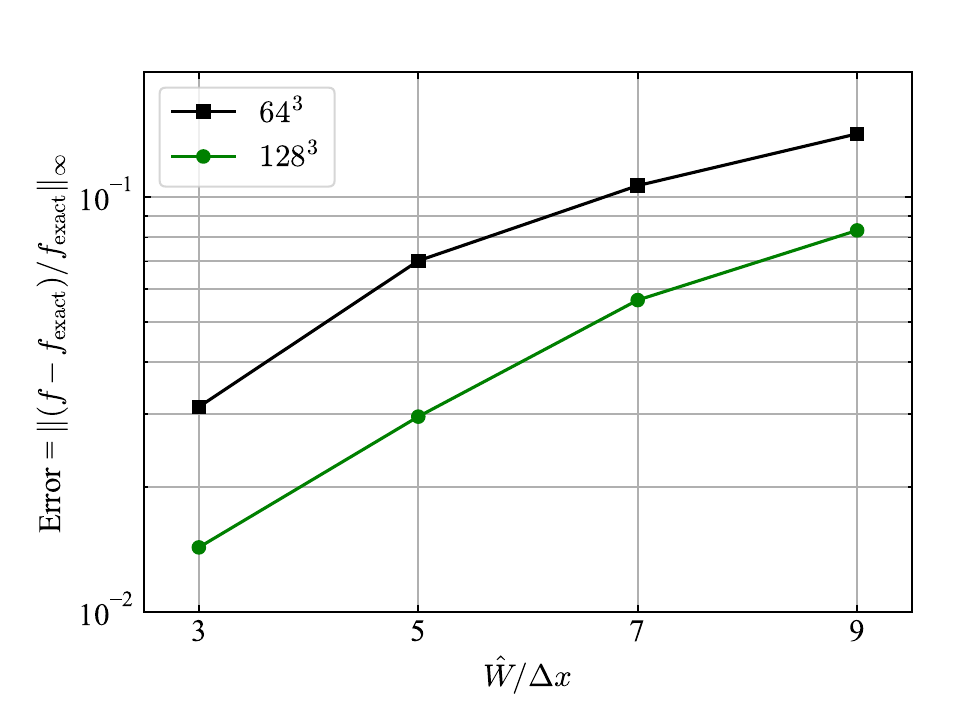}
    \caption{Errors in the surfactant concentration $f$ at $t = T$ for the $f$-type model using various delta function widths ($\hat{W} = 3 \Delta x$, $5 \Delta x$, $7 \Delta x$, and $9 \Delta x$) in the 3D vortical flow test described in Section~\ref{subsec:vortex3D}.}
    \label{fig:vortex3D_error_vs_width}
\end{figure}

Figure~\ref{fig:vortex3D_error_vs_width} shows the errors in the surfactant concentration for four delta function widths, $\hat{W} = 3 \Delta x$, $5 \Delta x$, $7 \Delta x$, and $9 \Delta x$, when using the $f$-type model.
The errors increase with increasing width for both the $64^3$ and $128^3$ cases.
As explained in the previous 2D case, this trend arises because an excessively wide delta function spreads a significant portion of the surfactant concentration $f_d$ far from the interface, and the velocity gradients near the interface distort the concentration profile.

\subsection{Surfactant-laden drop deformation in 2D shear flow}
\label{subsec:shear_flow_2D}

To demonstrate the applicability of the proposed approaches to two-phase flow simulations involving surfactants, we present a 2D simulation of a surfactant-laden drop deformed in a shear flow.  
A drop with radius $r_0 = 1~\mathrm{mm}$ is placed at the center of a computational domain $[-6~\mathrm{mm}, 6~\mathrm{mm}] \times [-2~\mathrm{mm}, 2~\mathrm{mm}]$.  
The left and right boundaries are periodic.  
The top and bottom boundaries are moving walls, with velocities prescribed as $(u, v) = (y, 0)$, resulting in a shear rate of $\dot{\gamma} = 1~\mathrm{s^{-1}}$.
The drop and the surrounding fluid have the same density and viscosity, $\rho = 10^3~\mathrm{kg/m^3}$ and $\mu = 10^{-3}~\mathrm{Pa \cdot s}$.
The surface tension coefficient of a clean interface is $\sigma_0 = 2 \times 10^{-6}~\mathrm{N/m}$.
Initially, the insoluble surfactant is uniformly distributed on the drop interface at a concentration of $f_0 = 10^{-6}~\mathrm{mol/m^2}$.  
The diffusion coefficient of the surfactant is $D = 10^{-7}~\mathrm{m^2/s}$.  
The effects of the surfactant on the fluid, i.e., surface tension reduction and the Marangoni effect, are incorporated into the surface tension terms of the Navier--Stokes equations as follows~\cite{ERIKTEIGEN2011375}:
\begin{equation}
    \frac{\partial (\rho \bm u)}{\partial t} + \nabla \cdot (\rho \bm u \bm u) = -\nabla p + \nabla \cdot \mu \left[ \nabla \bm u + \left( \nabla \bm u \right)^\top \right] + \sigma \kappa \nabla \phi +  \delta_\Gamma \nabla_\Gamma \sigma,
    \label{eq:NSeq}
\end{equation}
where $p$ is the pressure and $\kappa$ is the interface curvature.  
The surface tension coefficient $\sigma$ depends on the surfactant concentration $f$ as
\begin{equation}
    \sigma = \sigma_0 \left[1 + \beta \ln \left( 1 - \frac{f}{f_\infty} \right) \right],
\end{equation}
where $\beta = 0.3$ is the elasticity number and $f_\infty = 2 \times 10^{-6}~\mathrm{mol/m^2}$ is the maximum surfactant concentration.  
The last two terms in Eq.~(\ref{eq:NSeq}) represent the surface tension forces.  
The first term, $\sigma \kappa \nabla \phi$, corresponds to the continuum surface force (CSF) formulation~\cite{BRACKBILL_1992} of the normal force.  
The second term, $\delta_\Gamma \nabla_\Gamma \sigma$, represents the tangential (Marangoni) force, where $\nabla_\Gamma \sigma = (I - \bm n \bm n^\top) \nabla \sigma$ is the surface gradient of the surface tension coefficient and $I$ is the identity tensor.  
The dimensionless governing equations are provided in \ref{sec:dimensionless_governing_equation}.
The resolution is set to $600 \times 200$, and the time step is $\Delta t = 2 \times 10^{-5}~\mathrm{s}$.  
The parameter $\overline{D}$ is set equal to $D = 10^{-7}~\mathrm{m^2/s}$, and the $f$-type model with $\hat{W} = 5 \Delta x$ is used.
Gravity is neglected.

\begin{figure}
    \centering
    \includegraphics[width=0.7\linewidth]{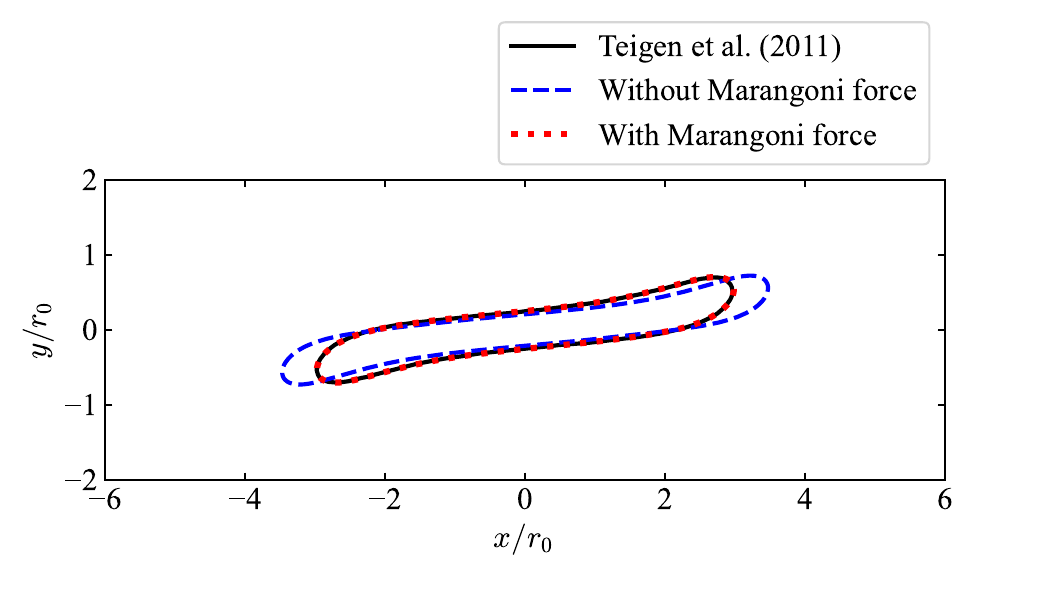}
    \caption{Drop shape at $t = 8~\mathrm{s}$ in the 2D shear flow test described in Section~\ref{subsec:shear_flow_2D}. The result reported by Teigen et al.~\cite{ERIKTEIGEN2011375} is used as a reference for the case with the Marangoni force.}
    \label{fig:shear_flow_2D_shape_t8}
\end{figure}

\begin{figure}
    \centering
    \includegraphics[width=0.7\linewidth]{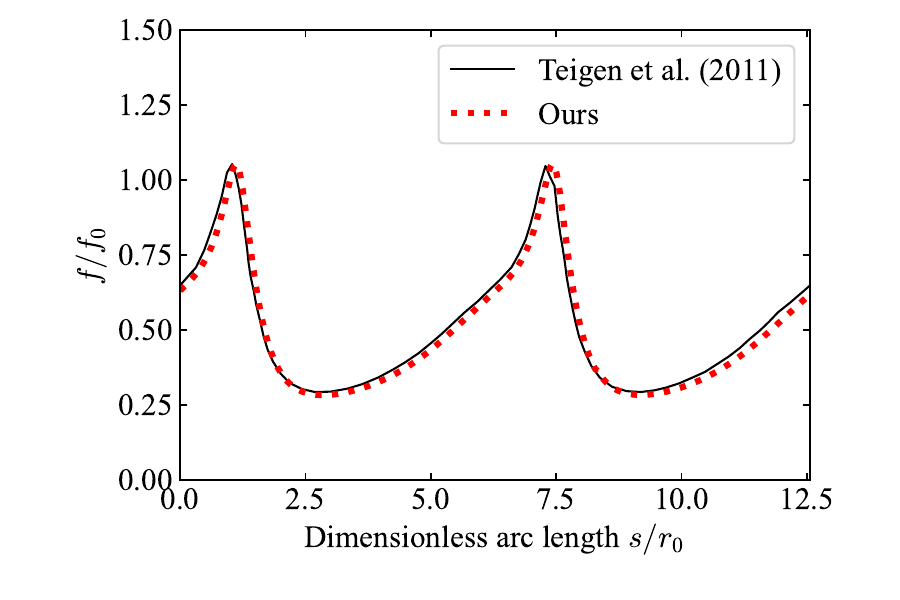}
    \caption{Surfactant concentration along the interface at $t = 8~\mathrm{s}$ in the 2D shear flow test (see Section~\ref{subsec:shear_flow_2D}). The arc length $s$ is measured from the point $y = 0$ on the right side of the drop and increases counterclockwise along the interface.}
    \label{fig:shear_flow_2D_concentration_t8}
\end{figure}

The dimensionless numbers corresponding to the above numerical settings are the same as those reported by Teigen et al.~\cite{ERIKTEIGEN2011375}:
\begin{align}
    \mathrm{Re} = \frac{\rho \dot{\gamma} r_0^2}{\mu} = 1, \quad
    \mathrm{Ca} = \frac{\mu \dot{\gamma} r_0}{\sigma_0} = 0.5, \quad
    \mathrm{Pe} = \frac{\dot{\gamma} r_0^2}{D} = 10, \quad
    \frac{f_0}{f_\infty} = 0.5, \quad
    \beta = 0.3,
\end{align}
where $\mathrm{Re}$, $\mathrm{Ca}$, and $\mathrm{Pe}$ denote the Reynolds, capillary, and Peclet numbers, respectively.  
Thus, our result can be directly compared with that of~\cite{ERIKTEIGEN2011375}.
Figure~\ref{fig:shear_flow_2D_shape_t8} shows the drop shape at $t = 8~\mathrm{s}$, with and without the Marangoni force in Eq.~(\ref{eq:NSeq}).
The drop deformation differs between the cases with and without the Marangoni force, indicating that this numerical test can assess whether the simulation correctly computes both surfactant transport and the resulting Marangoni force.
With the Marangoni force included, our result shows good agreement with that of Teigen et al.~\cite{ERIKTEIGEN2011375}. 
Moreover, Fig.~\ref{fig:shear_flow_2D_concentration_t8} compares the surfactant concentration along the interface with that reported by Teigen et al.~\cite{ERIKTEIGEN2011375}, showing good agreement.
Therefore, the proposed approaches are applicable to two-phase flow simulations involving surfactants.

\subsection{Limitation and proposed challenging benchmark}
\label{subsec:single_vortex_large}
Although numerous studies have proposed methods for interfacial surfactant transport, most have been validated only on test cases where achieving high accuracy is relatively straightforward. 
In this subsection, we present a challenging test case that remains difficult to solve accurately, even with the proposed approaches.
This test case serves as a valuable benchmark for evaluating and comparing various surfactant transport models developed in existing studies.

\begin{figure}
    \centering
    \includegraphics[width=\linewidth]{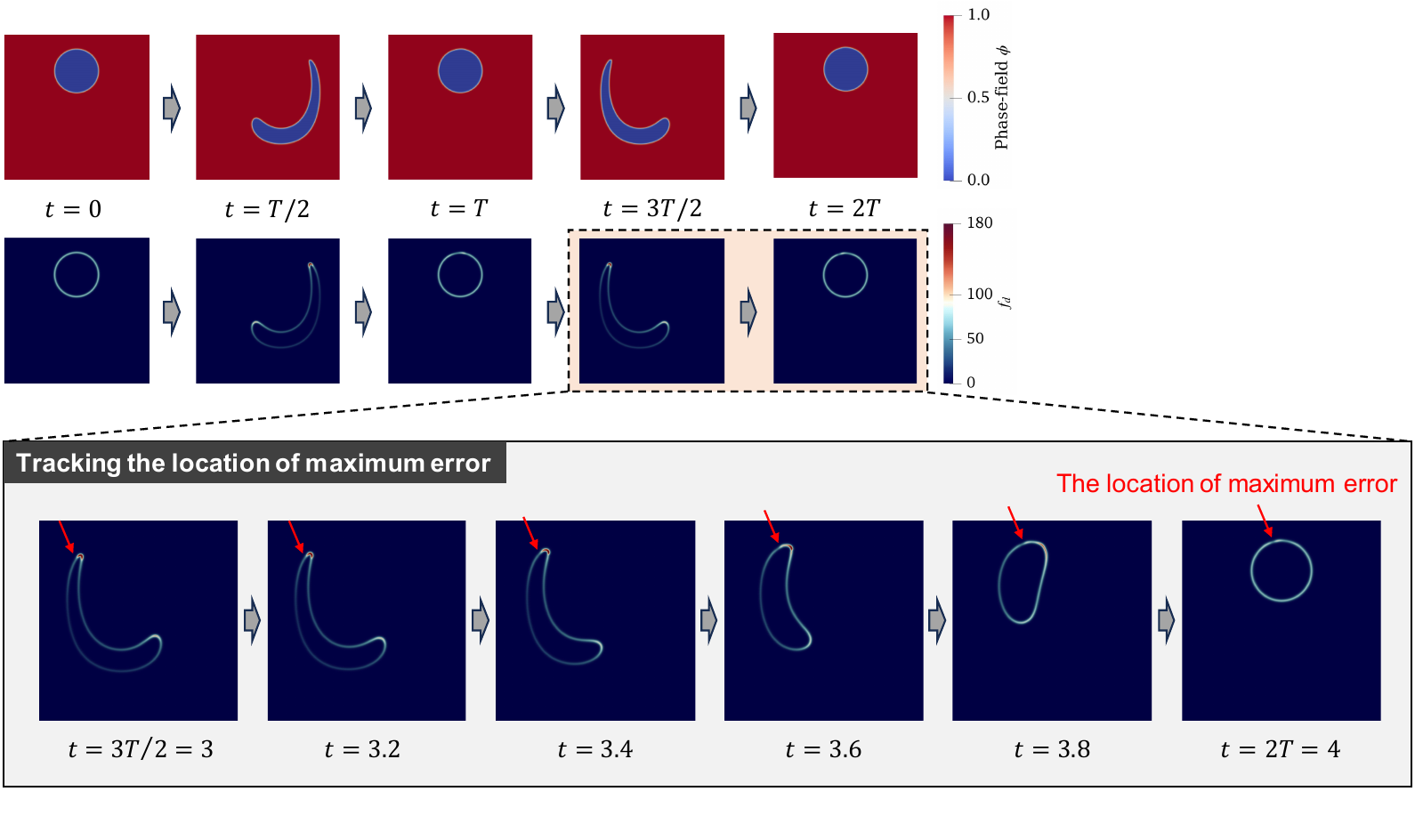}
    \caption{Interface deformation and surfactant concentration in the challenging benchmark test described in Section~\ref{subsec:single_vortex_large}. The deformation is intensified by increasing the vortex flow period from $T = 1$ to $T = 2$, compared to the setup in Section~\ref{subsec:single_vortex}. The bottom row shows the tracking of the location of the maximum error at $t = 2T$.}
    \label{fig:single_vortex_large_256_Delta03dx_f}
\end{figure}

\begin{figure}
    \centering
    \includegraphics[width=0.7\linewidth]{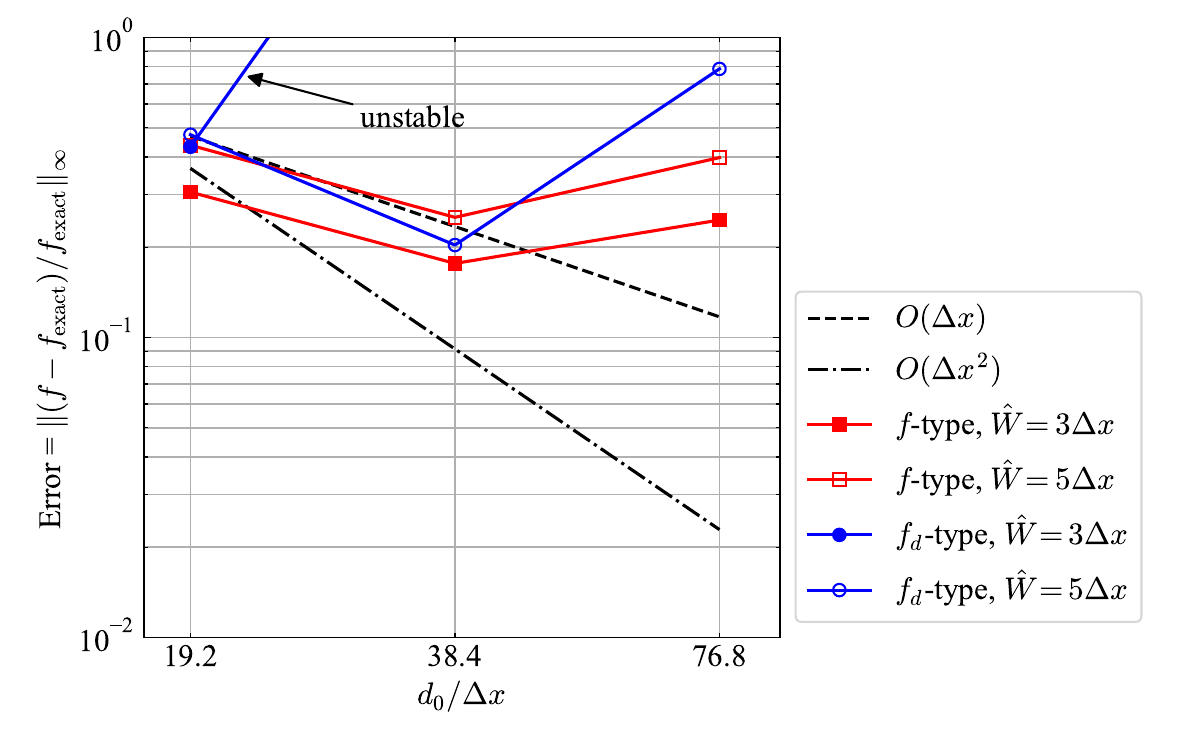}
    \caption{Surfactant concentration error at $t = 2T$ in the highly deforming benchmark case described in Section~\ref{subsec:single_vortex_large}.}
    \label{fig:single_vortex_large_convergence}
\end{figure}

\begin{figure}
    \centering
    \includegraphics[width=0.8\linewidth]{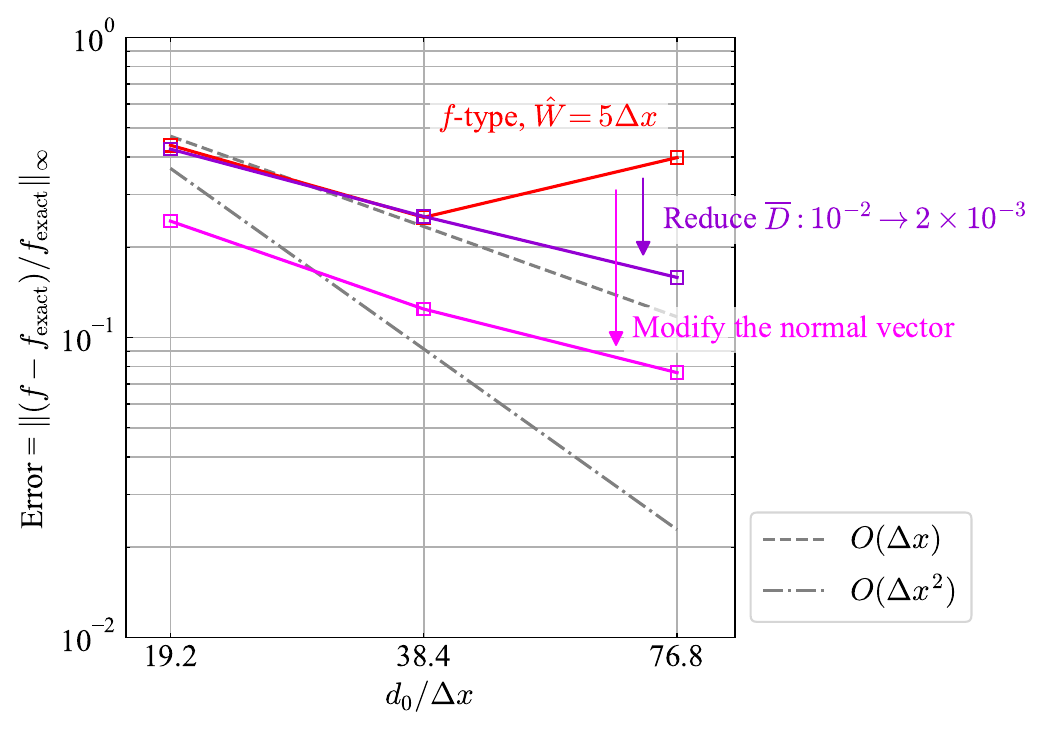}
    \caption{Improved accuracy of surfactant transport in the highly deforming benchmark case (Section~\ref{subsec:single_vortex_large}) is demonstrated for the $f$-type model with $\hat{W} = 5 \Delta x$. This improvement is achieved by two simple measures: (1) reducing $\overline{D}$ from $10^{-2}$ to $2 \times 10^{-3}$, and (2) modifying the normal vector where $|\nabla \psi|$ deviates from unity.
}
    \label{fig:single_vortex_large_convergence_improve}
\end{figure}

In this test, bubble deformation is intensified by increasing the parameter $T$ in Eq.~(\ref{eq:vortex2D_velocity}) from $T = 1$ to $T = 2$, based on the test case described in Section~\ref{subsec:single_vortex}.  
Figure~\ref{fig:single_vortex_large_256_Delta03dx_f} illustrates the interface deformation and corresponding surfactant concentration. 
Figure~\ref{fig:single_vortex_large_convergence} shows the results of the grid convergence study. 
As described in Sections~\ref{subsec:diffusion_in_uniform_flow}--\ref{subsec:vortex3D}, using a wider delta function ($\hat{W} = 5 \Delta x$) stabilizes the simulation for the $f_d$-type model.
For the $f$-type model, the simulation using the narrow delta function is stable and yields smaller errors than that using the wider delta function.
The narrow delta function reduces the distortion of the surfactant concentration profile caused by the velocity gradient near the interface, thereby improving accuracy. 
However, the error does not converge in any case.
In the bottom row of Figure~\ref{fig:single_vortex_large_256_Delta03dx_f}, the location of the maximum error at $t = 2T$ is highlighted and tracked with an arrow. 
This region corresponds to a tail-like structure at $t = 3T/2$. 
These results indicate that the accuracy of the surfactant transport deteriorates significantly in the sharp part of the interface. 
This degradation is primarily caused by the reduced accuracy of the computed normal vector in these regions.  
The following measures can be taken to mitigate this issue:
\begin{itemize}
\item \textbf{Reduce $\overline{D}$}.  
Reducing $\overline{D}$ helps mitigate the negative impact of inaccurate normal vectors on the accuracy of surfactant transport.  
Figure~\ref{fig:single_vortex_large_convergence_improve} shows that, for the $f$-type model with $\hat{W} = 5 \Delta x$, reducing $\overline{D}$ from $10^{-2}$ to $2 \times 10^{-3}$ improves the accuracy of surfactant transport.
\item \textbf{Modify the normal vector}.  
There may be various ways to modify the normal vector to improve its accuracy.  
One example is to set $\bm{n} = \bm{0}$ at cell centers where $|\nabla \psi|$ deviates from unity, specifically when $| \nabla \psi | < 0.99$ or $1.01 < | \nabla \psi |$.
Figure~\ref{fig:single_vortex_large_convergence_improve} shows that this strategy improves the accuracy of surfactant transport for the $f$-type model with $\hat{W} = 5 \Delta x$.
\end{itemize}
Although these measures are straightforward to implement, the resulting convergence rates remain below first order.
Because sharp structures frequently occur in two-phase flows with surfactants, the ability to solve them accurately is essential. 
Therefore, this test case highlights the need for further development of surfactant transport models capable of handling these challenging conditions.

\clearpage

\section{Conclusions}
\label{sec:conclusion}

This study focused on enhancing the accuracy of interfacial transport models for insoluble surfactants using the diffuse-interface method in two-phase flows.  
Our main contributions are as follows:
\begin{itemize}
    \item We presented two generalized transport models ($f_d$-type and $f$-type models) and proposed two approaches to improve their accuracy (Section~\ref{sec:approaches}).
    \begin{itemize}
        \item The first approach is adopting the formulation that avoids spatial derivatives of variables with sharp gradients; that is, using the $f$-type model instead of the $f_d$-type model (Section~\ref{subsec:approach1}).
        Although the two models are mathematically equivalent at the equilibrium of the phase-field variable, their numerical accuracies differ significantly. 
        This difference has not been emphasized in previous studies.
        \item The second approach is decoupling the width of the delta function from the interface width (Section~\ref{subsec:approach2}).
        We introduced a method that allows the width of the delta function to be specified independently of the interface width, enabling improved surfactant transport accuracy without sacrificing the interface-capturing quality or reducing the allowable time step. 
        This strategy has not been systematically explored in previous diffuse-interface studies.
    \end{itemize}
    These approaches are simple and practical in that they do not significantly increase computational cost, implementation complexity, or degrade the accuracy of interface capturing.  
    Moreover, they preserve the discrete conservation of both fluid and surfactant mass.
    \item We validated the proposed approaches through a series of numerical tests (Section~\ref{sec:numerical_tests}).
    \begin{itemize}
        \item We confirmed that both proposed approaches improve accuracy and stability in simulations involving a wide range of advection-to-diffusion ratios and deforming interfaces.
        The approaches are effective in both 2D and 3D. 
        Furthermore, we demonstrated their applicability to surfactant-laden two-phase flows by simulating a deformed drop in shear flow. 
        The simulation accurately captured both surfactant transport and the resulting Marangoni force.
        \item We clarified how the width of the delta function affects the accuracy.
        The width of the delta function should be set between approximately $3 \Delta x$ and $6 \Delta x$.
        If the delta function is extremely narrow, the computational grid cannot adequately resolve its sharp profile, leading to significant errors in both the advection and diffusion terms.
        In contrast, if the delta function is extremely wide, a substantial portion of the surfactant concentration is distributed away from the interface, degrading the accuracy of interfacial diffusion and distorting the concentration profile due to the velocity gradient near the interface.
    \end{itemize}
    \item We introduced a challenging benchmark test that is difficult to solve accurately~(Section~\ref{subsec:single_vortex_large}). In this test, the interface undergoes severe deformation and develops a sharp profile, resulting in significant errors in the normal vector. As a result, the surfactant concentration error does not converge with grid refinement. Although this problem is critical for a wide range of surfactant-laden two-phase flow simulations, it has largely been overlooked. Therefore, this benchmark can serve as a valuable reference for evaluating and comparing surfactant transport models proposed in previous studies.
\end{itemize}
We believe that this study will support future efforts to develop more accurate and robust models for interfacial surfactant transport.

\section*{Declaration of generative AI and AI-assisted technologies in the writing process}

During the preparation of this work the authors used ChatGPT in order to improve the readability and language of the manuscript. 
After using this tool, the authors reviewed and edited the content as needed and take full responsibility for the content of the published article.

\section*{Acknowledgements}

This work was supported by JSPS KAKENHI Grant Number 22H03770, 22K14178, and 25K07582, and Japan Science and Technology Agency (JST) as part of Adopting Sustainable Partnerships for Innovative Research Ecosystem (ASPIRE), Grant Number JPMJAP2407. 
This work was also supported by JST SPRING, Japan Grant Number JPMJSP2180.
This study was carried out using the TSUBAME4.0 supercomputer at Institute of Science Tokyo.

\appendix

\section{Equivalence between $f_d$-type and $f$-type models}
\label{sec:equivalence_f_and_fd}

We demonstrate the equivalence between the $f_d$-type model [Eq.~\ref{eq:fd_type}] and the $f$-type model [Eq.~(\ref{eq:f_type})] under the assumption that the phase-field variable $\phi$ maintains its equilibrium profile.
Under this assumption, $\nabla \phi$ can be expressed as
\begin{align}
    \nabla \phi 
    &= \nabla \frac{1}{2} \left[ 1 + \tanh \left( \frac{\psi}{2 \epsilon} \right) \right] \nonumber \\
    &= \frac{1}{2} \left[ 1 - \tanh^2 \left( \frac{\psi}{2 \epsilon} \right) \right] \frac{\nabla \psi}{2 \epsilon} \nonumber \\
    &= \frac{1}{2} \left[ 1 + \tanh \left( \frac{\psi}{2 \epsilon} \right) \right] \frac{1}{2} \left[ 1 - \tanh \left( \frac{\psi}{2 \epsilon} \right) \right] \frac{\nabla \psi}{\epsilon} \nonumber \\
    &= \frac{\phi (1 - \phi)}{\epsilon} \nabla \psi \nonumber \\
    &= \delta_\Gamma \bm n.
\end{align}
Accordingly, $\nabla \delta_\Gamma$ is given by
\begin{align}
    \nabla \delta_\Gamma 
    &= \nabla \frac{\phi (1 - \phi)}{\epsilon} \nonumber \\
    &= \frac{(1 - 2 \phi) \nabla \phi}{\epsilon} \nonumber \\
    &= \frac{2 (0.5 - \phi) \bm n \delta_\Gamma}{\epsilon}.
\end{align}
Using this relation, the $f_d$-type model can be rewritten as
\begin{align}
    \frac{\partial f_d}{\partial t} + \nabla \cdot (\bm u f_d) 
    &= \nabla \cdot D \left[ \nabla f_d - \frac{2 (0.5 - \phi) \bm n f_d}{\epsilon} \right] 
    + \nabla \cdot \overline{D} \left[ \bm n \bm n^\top \nabla f_d - \frac{2 (0.5 - \phi) \bm n f_d}{\epsilon} \right] \nonumber \\
    &= \nabla \cdot D \left[ \nabla (f \delta_\Gamma) - f \frac{2 (0.5 - \phi) \bm n \delta_\Gamma}{\epsilon} \right] 
    + \nabla \cdot \left\{ \overline{D} \bm n \bm n^\top \left[ \nabla (f \delta_\Gamma) - f \frac{2 (0.5 - \phi) \bm n \delta_\Gamma}{\epsilon} \right] \right\} \nonumber \\
    &= \nabla \cdot D \left[ \nabla (f \delta_\Gamma) - f \nabla \delta_\Gamma \right] 
    + \nabla \cdot \left\{ \overline{D} \bm n \bm n^\top \left[ \nabla (f \delta_\Gamma) - f \nabla \delta_\Gamma \right] \right\} \nonumber \\
    &= \nabla \cdot \left( D \delta_\Gamma \nabla f \right) 
    + \nabla \cdot \left( \overline{D} \delta_\Gamma \bm n \bm n^\top \nabla f \right).
\end{align}
Therefore, the $f_d$-type model is equivalent to the $f$-type model.

\section{Detailed implementation}
\label{sec:detailed_implementation}

We present a detailed version of the implementation described in Section~\ref{sec:overview_implementation}. 
For simplicity, we focus on the 2D case and assume the use of a first-order explicit Euler time integration. 
The extension to 3D and higher-order time integration schemes is straightforward. 
A cell center is denoted by the index pair $(i, j)$, with the corresponding left cell face located at $(i - 1/2, j)$ and the bottom cell face at $(i, j - 1/2)$. 
Linear interpolation at cell faces from the adjacent cell centers is denoted by $\langle q \rangle_{i - 1/2, j} = (q_{i - 1, j} + q_{i, j}) / 2$ or $\langle q \rangle_{i, j - 1/2} = (q_{i, j - 1} + q_{i, j}) / 2$.

The numerical procedure from time step $k$ to $k + 1$ is as follows:
\begin{enumerate}
    \item If $k \equiv 0 \pmod{20}$, reinitialize the level-set function $\psi^k$.
    \begin{itemize}
        \item Compute $\psi_0$ by overwriting $\psi^k$ near the interface $(0.1 < \phi^k < 0.9)$:
        \begin{equation*}
            \psi_{0,i,j} =
            \begin{cases}
                \epsilon \ln \left( \frac{ {}^c\phi_{i,j}^k + 10^{-100}}{1 - {}^c\phi_{i,j}^k + 10^{-100}} \right) & \text{if $0.1 < \phi_{i,j}^k < 0.9$,} \\
                \psi_{i,j}^k & \text{otherwise,}
            \end{cases}
        \end{equation*}
        where ${}^c\phi = \max(\min(\phi, 1), 0)$.
        \item Solve Eq.~(\ref{eq:reinit_ls}) for 20 iterations with $\psi_0$ as the initial condition, using the second-order Runge-Kutta scheme~\cite{SHU1988439}.
        A subcell fix~\cite{MIN20102764} is employed to prevent the interface from moving during this iteration.
        In cells whose centers are located within $0.01 \Delta x$ from the interface, we set $\psi_{i,j}^k = 0$, as described in~\cite{MIN2007300}.
        \item Update $\hat{\phi}_{i,j}^k = \frac{1}{2} \left[ 1 + \tanh \left( \frac{\psi_{i,j}^k}{2 \hat{\epsilon}} \right) \right]$.
    \end{itemize}
    \item Compute the phase-field variable $\phi^{k+1}$ using the ACDI method [Eq.~(\ref{eq:acdi})]. 
    The finite volume method is employed, as detailed in~\cite{JAIN2022111529}.
    \item Compute the level-set function $\psi^{k+1}$ by solving the advection equation [Eq.~(\ref{eq:ls_advection})]:  
\begin{itemize}
    \item Compute the cell-centered velocity as $\bm u_{i,j}^k = \left( \frac{u_{i - 1/2, j}^k + u_{i + 1/2, j}^k}{2}, \frac{v_{i, j - 1/2}^k + v_{i, j + 1/2}^k}{2} \right)$.
    \item Compute $\nabla \psi_{i,j}^k$ using the WENO scheme~\cite{Jiang_2000}.
    \item Update the level-set function by $\psi_{i,j}^{k+1} = \psi_{i,j}^k - \Delta t \left( \bm u_{i,j}^k \cdot \nabla \psi_{i,j}^k \right)$.
\end{itemize}
    \item Compute $\hat{\phi}_{i,j}^{k+1} = \frac{1}{2} \left[ 1 + \tanh \left(\frac{\psi_{i,j}^{k+1}}{2 \hat{\epsilon}}\right) \right]$ and $\hat{\delta_\Gamma}_{i,j}^{k+1} = \frac{\hat{\phi}_{i,j}^{k+1} \left(1 - \hat{\phi}_{i,j}^{k+1}\right)}{\hat{\epsilon}}$.
    \item Compute the surfactant concentration $f_d^{k+1}$ and $f^{k+1}$:
    \begin{itemize}
        \item Compute $\nabla f_d^k$ at cell faces for the $f_d$-type model [Eq.~(\ref{eq:fd_type})], or $\nabla f^k$ at cell faces for the $f$-type model [Eq.~(\ref{eq:f_type})]:
        \begin{align*}
            \nabla f_{d,i-1/2,j}^k &= \left( \frac{f_{d,i,j}^k - f_{d,i-1,j}^k}{\Delta x}, \frac{\left\langle f_d^k \right\rangle _{i-1/2,j+1} - \left\langle f_d^k \right\rangle _{i-1/2,j-1}}{2 \Delta y}\right), \\
            \nabla f_{d,i,j-1/2}^k &= \left( \frac{\left\langle f_d^k \right\rangle _{i+1,j-1/2} - \left\langle f_d^k \right\rangle _{i-1,j-1/2}}{2\Delta x}, \frac{f_{d,i,j}^k - f_{d,i,j-1}^k}{\Delta y} \right). \\
            \nabla f_{i-1/2,j}^k &= \left( \frac{f_{i,j}^k - f_{i-1,j}^k}{\Delta x}, \frac{\left\langle f^k \right\rangle _{i-1/2,j+1} - \left\langle f^k \right\rangle _{i-1/2,j-1}}{2 \Delta y}\right), \\
            \nabla f_{i,j-1/2}^k &= \left( \frac{\left\langle f^k \right\rangle _{i+1,j-1/2} - \left\langle f^k \right\rangle _{i-1,j-1/2}}{2\Delta x}, \frac{f_{i,j}^k - f_{i,j-1}^k}{\Delta y} \right).
        \end{align*}
        \item For the $f_d$-type model, compute the flux $\bm a = \left( a_x, a_y \right)$ at cell faces:
        \begin{align*}
            a_{x, i-1/2, j} = 
            &- \left\langle f_d^k \right\rangle _{i-1/2,j} u_{i - 1/2, j}^k \\
            &+ D \left[ \left(\frac{\partial f_d^k}{\partial x}\right)_{i-1/2,j} - \frac{2}{\hat \epsilon} \left(0.5 - \left\langle \hat{\phi}^k \right\rangle _{i-1/2,j} \right) \left\langle n_x^k \right\rangle _{i-1/2,j} \left\langle f_d^k \right\rangle _{i-1/2,j}\right] \\
            &+ \overline{D} \left[ \left\langle n_x^k \right\rangle _{i-1/2,j} \left\langle \bm n^k \right\rangle _{i-1/2,j} \cdot \nabla f_{d,i-1/2,j}^k - \frac{2}{\hat \epsilon} \left(0.5 - \left\langle \hat{\phi}^k \right\rangle _{i-1/2,j} \right) \left\langle n_x^k \right\rangle _{i-1/2,j} \left\langle f_d^k \right\rangle _{i-1/2,j}\right], \\
            a_{y, i, j-1/2} = 
            &- \left\langle f_d^k \right\rangle _{i,j-1/2} v_{i, j - 1/2}^k \\
            &+ D \left[ \left(\frac{\partial f_d^k}{\partial y}\right)_{i,j-1/2} - \frac{2}{\hat \epsilon} \left(0.5 - \left\langle \hat{\phi}^k \right\rangle _{i,j-1/2}\right) \left\langle n_y^k \right\rangle _{i,j-1/2} \left\langle f_d^k \right\rangle _{i,j-1/2}\right] \\
            &+\overline{D} \left[ \left\langle n_y^k \right\rangle _{i,j-1/2} \left\langle \bm n^k \right\rangle _{i,j-1/2} \cdot \nabla f_{d,i,j-1/2}^k - \frac{2}{\hat \epsilon} \left(0.5 - \left\langle \hat{\phi}^k \right\rangle _{i,j-1/2}\right) \left\langle n_y^k \right\rangle _{i,j-1/2} \left\langle f_d^k \right\rangle _{i,j-1/2}\right].
        \end{align*}
        \item For the $f$-type model, compute the flux $\bm a$ at cell faces:
        \begin{align*}
            a_{x, i-1/2, j} = 
            &- \left\langle f_d^k \right\rangle _{i-1/2,j} u_{i - 1/2, j}^k \\
            &+ D \frac{ \left\langle \hat{\phi}^k \right\rangle _{i-1/2,j} \left( 1 - \left\langle \hat{\phi}^k \right\rangle _{i-1/2,j} \right)}{\hat \epsilon} \left(\frac{\partial f^k}{\partial x}\right)_{i-1/2,j} \\ 
            &+ \overline{D} \frac{ \left\langle \hat{\phi}^k \right\rangle _{i-1/2,j} \left( 1 - \left\langle \hat{\phi}^k \right\rangle _{i-1/2,j} \right)}{\hat \epsilon} \left\langle n_x^k \right\rangle _{i-1/2,j} \left\langle \bm n^k \right\rangle _{i-1/2,j} \cdot \nabla f_{i-1/2,j}^k, \\
            a_{y, i, j-1/2} = 
            &- \left\langle f_d^k \right\rangle _{i,j-1/2} v_{i, j - 1/2}^k \\
            &+ D \frac{\left\langle \hat{\phi}^k \right\rangle _{i,j-1/2} \left( 1 - \left\langle \hat{\phi}^k \right\rangle _{i,j-1/2} \right)}{\hat \epsilon} \left(\frac{\partial f^k}{\partial y}\right)_{i,j-1/2} \\
            &+ \overline{D} \frac{ \left\langle \hat{\phi}^k \right\rangle _{i,j-1/2} \left( 1 - \left\langle \hat{\phi}^k \right\rangle _{i,j-1/2} \right)}{\hat \epsilon} \left\langle n_y^k \right\rangle _{i,j-1/2} \left\langle \bm n^k \right\rangle _{i,j-1/2} \cdot \nabla f_{i,j-1/2}^k.
        \end{align*}
        \item Compute $f_{d,i,j}^{k+1} = f_{d,i,j}^k + \Delta t \left( \frac{a_{x,i+1/2,j} - a_{x,i-1/2,j}}{\Delta x} + \frac{a_{y,i,j+1/2} - a_{y,i,j-1/2}}{\Delta y} \right)$.
        \item Compute $f_{i,j}^{k+1} = \frac{f_{d,i,j}^{k+1}}{\hat{\delta_\Gamma}_{i,j}^{k+1} + 10^{-5}}$.
    \end{itemize}
    \item Compute the normal vector $\bm n^{k+1}$ at cell centers using Eq.~(\ref{eq:normal_from_ls}):
    \begin{itemize}
        \item Compute $\nabla \psi_{i,j}^{k+1} = \left( \frac{\psi_{i+1, j}^{k+1} - \psi_{i-1, j}^{k+1}}{2 \Delta x}, \frac{\psi_{i,j+1}^{k+1} - \psi_{i,j-1}^{k+1}}{2 \Delta y} \right)$.
        \item Compute the normal vector $\bm n_{i,j}^{k+1} = \frac{\nabla \psi_{i,j}^{k+1}}{\left|\nabla \psi_{i,j}^{k+1}\right| + 10^{-100}}$
    \end{itemize} 
    \item Update $\bm{u}^{k+1}$ by applying a prescribed velocity field or solving the fluid equations.
    \item Update the parameter $\gamma^{k+1} = 1.1 \left| \bm{u}^{k+1} \right|_\mathrm{max}$ in the ACDI model [Eq.~(\ref{eq:acdi})].
    \item Advance time by setting $t = t + \Delta t$, and return to Step 1 with $k = k + 1$.
\end{enumerate}

\section{Dimensionless governing equations}
\label{sec:dimensionless_governing_equation}

We present the dimensionless governing equations for the simulation of surfactant-laden drop deformation in a 2D shear flow (see Section~\ref{subsec:shear_flow_2D}).
The characteristic scales are chosen as $L = r_0$ for length, $T = \dot{\gamma}^{-1}$ for time, $U = \dot{\gamma} r_0$ for velocity, $\rho U^2 = \rho \dot{\gamma}^2 r_0^2$ for pressure, $\sigma_0$ for surface tension coefficient, and the initial concentration $f_0$ for the surfactant concentration.
Thus, the Navier--Stokes equation in Eq.~(\ref{eq:NSeq}) is non-dimensionalized as
\begin{equation}
    \frac{\partial \bm u^\ast}{\partial t^\ast} + \nabla^\ast \cdot (\bm u^\ast \bm u^\ast) = -\nabla^\ast p^\ast + \frac{1}{\mathrm{Re}} \nabla^\ast \cdot \left[ \nabla^\ast \bm u^\ast + \left( \nabla^\ast \bm u^\ast \right)^\top \right] + \frac{1}{\mathrm{Re}\,\mathrm{Ca}} \left( \sigma^\ast \kappa^\ast \nabla^\ast \phi + \delta_\Gamma^\ast \nabla_\Gamma^\ast \sigma^\ast \right),
\end{equation}
where the superscript $\ast$ denotes dimensionless variables.
Here, $\mathrm{Re} = \rho U L / \mu = \rho \dot{\gamma} r_0^2 / \mu$ is the Reynolds number, and $\mathrm{Ca} = \mu U / \sigma_0 = \mu \dot{\gamma} r_0 / \sigma_0$ is the capillary number.
The dimensionless surface tension coefficient $\sigma^\ast$ is given by
\begin{equation}
    \sigma^\ast = \frac{\sigma}{\sigma_0} = 1 + \beta \ln \left( 1 - \frac{f_0}{f_\infty} f^\ast \right).
\end{equation}
The surfactant transport equation in Eq.~(\ref{eq:f_type}) is non-dimensionalized as
\begin{equation}
    \frac{\partial f_d^\ast}{\partial t^\ast} + \nabla^\ast \cdot (\bm u^\ast f_d^\ast) = \frac{1}{\mathrm{Pe}} \nabla^\ast \cdot \left( \delta_\Gamma^\ast \nabla^\ast f^\ast \right) + \frac{1}{\overline{\mathrm{Pe}}} \nabla^\ast \cdot \left[ \delta_\Gamma^\ast \bm n \bm n^\top \nabla^\ast f^\ast \right],
\end{equation}
where $\mathrm{Pe} = U L / D = \dot{\gamma} r_0^2 / D$ and $\overline{\mathrm{Pe}} = U L / \overline{D} = \dot{\gamma} r_0^2 / \overline{D}$ are the Peclet numbers.
These equations are solved together with the continuity equation, e.g., $\nabla^\ast \cdot \bm u^\ast = 0$ for the incompressible scheme.

\bibliographystyle{elsarticle-num} 
\bibliography{reference}

\end{document}